\documentclass[a4paper, 12pt]{article}

\usepackage[margin=1.2in]{geometry}
\usepackage[utf8]{inputenc}
\usepackage[style=apa]{biblatex}
\usepackage{amsfonts,amsmath,amsthm,mathtools,MnSymbol}
\usepackage{paralist}
\usepackage[ruled,linesnumbered]{algorithm2e}
\usepackage{xcolor}
\usepackage{tcolorbox}
\usepackage[hang,flushmargin]{footmisc}
\usepackage{caption}
\usepackage{subcaption}
\usepackage{setspace}
\usepackage{authblk}
\usepackage{booktabs}

\theoremstyle{definition}

\SetKwInput{KwInput}{Input}
\SetKwInput{KwOutput}{Output}
\SetKw{KwFrom}{from}
\SetKw{KwAnd}{and}
\SetKw{KwOr}{or}
\SetKwComment{KwComment}{/* }{ */}
\SetKwComment{KwInline}{}{}

\let\oldnl\nl
\newcommand{\nonl}{\renewcommand{\nl}{\let\nl\oldnl}}
\makeatletter
\newcommand{\algorithmfootnote}[2][\footnotesize]{%
  \let\old@algocf@finish\@algocf@finish%
  \def\@algocf@finish{\old@algocf@finish%
    \leavevmode\rlap{\begin{minipage}{\linewidth}
    #1#2
    \end{minipage}}%
  }%
}
\makeatother

\DeclareMathOperator{\Var}{Var}
\DeclareMathOperator{\sign}{sign}
\DeclareMathOperator{\SRSF}{SRSF}
\DeclareMathOperator{\expit}{expit}
\DeclareMathOperator{\closure}{closure}
\DeclareMathOperator{\argmin}{argmin}
\DeclareMathOperator{\arginf}{arginf}
\newcommand{\diff}{\mathop{}\!{d}}
\DeclareMathOperator{\NN}{NN}

\DeclareMathOperator{\ToWarp}{ToWarp}
\DeclareMathOperator{\KarcherMean}{KarcherMean}

\DeclareMathOperator{\spl}{spl}

\DeclareMathOperator{\ext}{ext}

\DeclareMathOperator{\Scale}{Scale}

\title{Joint Alignment of Multivariate Quasi-Periodic Functional Data Using Deep Learning}
\author[1]{Vi Thanh Pham}
\author[2]{Jonas Bille Nielsen}
\author[2,3]{Klaus Fuglsang Kofoed}
\author[4]{Jørgen Tobias Kühl}
\author[1]{Andreas Kryger Jensen}
\affil[1]{Section of Biostatistics, Department of Public Health, Faculty of Health and Medical Sciences, University of Copenhagen}
\affil[2]{Department of Cardiology and Radiology, Copenhagen University Hospital}
\affil[3]{Department of Clinical Medicine, Faculty of Health and Medical Sciences, 
University of Copenhagen}
\affil[4]{Department of Cardiology, Zealand University Hospital}
\date{}

\begin{document}
\maketitle

\section*{Abstract}
The joint alignment of multivariate functional data plays an important role in various fields such as signal processing, neuroscience and medicine, including the statistical analysis of data from wearable devices. Traditional methods often ignore the phase variability and instead focus on the variability in the observed amplitude. We present a novel method for joint alignment of multivariate quasi-periodic functions using deep neural networks, decomposing, but retaining all the information in the data by preserving both phase and amplitude variability. Our proposed neural network uses a special activation of the output that builds on the unit simplex transformation, and we utilize a loss function based on the Fisher-Rao metric to train our model. Furthermore, our method is unsupervised and can provide an optimal common template function as well as subject-specific templates. We demonstrate our method on two simulated datasets and one real example, comprising data from 12-lead 10s electrocardiogram recordings.

\begin{center}
\textbf{Keywords}: Deep Learning; Elastic Phase-Amplitude Separation; Functional Data; Joint Multivariate Alignment; Multiscale Time Warping; Quasi-Periodic Functions.
\end{center}

\section{Introduction}
Statistical analysis of functional data is becoming increasingly essential in the biological and medical research as technologies allow for measuring subjects over long time durations with potential high-frequency sampling times. When applying statistical inference to functional data, it would be advantageous to align the functions, so that the positions of corresponding peaks and valleys are the same across the data set. Failure to align the data correctly can lead to inefficiency of basic statistical summaries like averages, leading to poor estimates of population mean functions that are not representative of the data. In reality, many commonly used functional data analysis techniques present inferior performance when confronted with phase variation \parencite{marronFunctionalDataAnalysis2015}. Furthermore, alignment can also improve prediction accuracy when classification is the goal \parencite{tuckerGenerativeModelsFunctional2013}. In practice, functional data is often misaligned and contains variability along both the $x$ axis, called the \textit{phase} variability, and the $y$ axis, called the \textit{amplitude} variability. The process of decomposing the overall variability in the data into these two components is termed elastic phase-amplitude separation and represents the joint alignment of multiple functions. In this paper, we focus on multivariate quasi-periodic functions. We call a function $f$ multivariate if $f(t)=\left(f_1(t),f_2(t),\dots,f_J(t)\right)\in\mathbb{R}^J$, and following \textcite{bouchehamMatchingQuasiperiodicTime2008}, we call a function quasi-periodic if it is a concatenation of similar patterns or pseudo-periods. 

Examples of such quasi-periodic functional data are continuous glucose monitoring (CGM) data \parencite{klonoffContinuousGlucoseMonitoring2005}, which provide information about levels of blood glucose through wearable devices \parencite{mcdonnellRegistration24hourAccelerometric2022}, and electrocardiogram (ECG) recordings \parencite{greggWhatElectrocardiograph2008}, which measure the electrical activity of the heart. We aim to study ECGs as they are regularly used together with other tests to diagnose and monitor diseases affecting the cardiovascular system, as well as to inspect symptoms of possible heart conditions. ECGs can assist with detecting arrhythmias and can be used over time to monitor a person with an existing diagnosis, or a person taking medication that can have an influence on the heart. These measurements taken over time through multiple leads placed on the body can be regarded as multivariate quasi-periodic functions, where the periods are the heartbeats. However, these periods are not exactly the same across the whole recording, and depend on other physiological activities at the moment, turning the data into quasi-periodic.

One can perform statistical analyses to align and compare these observations using tools such as the $\mathrm{L}^2$ distance to find the cross-sectional mean and variance. In addition to these summary measures, it is of great interest to model the variability in the data, and due to the high dimensionality of this data, we would like to analyze them using neural networks. In the example of the ECG data, we might want to align the peaks and segments to extract a template heartbeat while also recording the relative positions of these features.

In this paper, to jointly (simultaneously) model multiple time warping functions and to align multivariate quasi-periodic functions along the $x$ axis, we modify and extend the elastic phase-amplitude separation algorithm for univariate functions from \textcite{srivastavaFunctionalShapeData2016} and introduce the algorithm for the Joint Alignment of Multivariate quasi-periodic functional data using Deep learning (DeepJAM). Importantly, DeepJAM provides us with warping functions, one for each observed function, that contain all the information about the phase variability in the data. In the case of multivariate functional data, all the dimensions share the same warping function, because they are all observed on the same sample. Because our approach is unsupervised, we also utilize nice geometric properties of transformed warping functions to extract a unique multivariate functional template. Subsequently, we can use the DeepJAM neural network to easily align new data to this template. Furthermore, DeepJAM can be used as an integrated part of other neural networks in an end-to-end analysis as opposed to two-step modelling, where alignment is only treated as a pre-processing step.

\subsection{Related work}
A standard approach for alignment of functions is called landmark registration \parencite[chap.~7]{ramsayFunctionalDataAnalysis2005}. Landmarks are typically distinctive features of a function, such as minima, maxima, and zero-crossings of the function itself, or its derivatives. The challenge with landmark registration is that it only focuses on the alignment of specific points. An extension to the landmark registration is template registration \parencite[chap.~8.2]{srivastavaFunctionalShapeData2016}, which aligns whole functions to a specific template. \textcite{kneipCombiningRegistrationFitting2008} suggest aligning functions to the functional principal components instead, with the idea that aligned functions can be represented as the sum of a mean function and a linear combination of a few principal components. This is done through an iterative process of simultaneous estimation of the mean function, warping functions, functional principal components and their scores. An alternative approach also based on functional principal component analysis (FPCA) is presented in \textcite[chap.~8.8]{srivastavaFunctionalShapeData2016}.

Dynamic Time Warping \parencite[DTW,][chap.~4]{mullerInformationRetrievalMusic2007} is another well-known technique for finding an optimal alignment between two time series. Building on DTW, \textcite{boulnemourQPDTWUpgradingDynamic2018} proposed a combination with the shape exchange algorithm, which should be more suitable for handling quasi-periodic time series. However, in applications, we often need joint alignment of multiple functions, because we are interested in patterns regarding the whole data, such as population mean and variance. This could be done, for example, by finding a representative sample in the data, or by constructing a representative template and then using DTW to align the functions to this template. Nonetheless, DTW does not provide us with this template. \textcite{srivastavaFunctionalShapeData2016} proposed a method to solve this problem using dynamic programming. Regardless of whether the objective is to achieve pairwise or groupwise alignment, the purpose of the above methods using dynamic programming is mainly to serve as a pre-processing step. In the case when we require an end-to-end analysis using neural networks, we need the alignment approach itself to be based on neural networks, or at least another gradient based approach.

In the deep learning universe, the main focus has been on time warping invariant neural networks \parencite{sunTimeWarpingInvariant1992}, convolutional neural networks \parencite{lecunDeepLearning2015} that can be invariant to shifts in the input data, and recurrent neural networks \parencite{tallecCanRecurrentNeural2018}. The challenge with using neural networks that are invariant to time warping or shifts is that they mostly focus on amplitude variability. This may or may not be relevant depending on the specific task. Disregarding the phase variability could result in loss of information and inadequate generative models, while using models for both amplitude and phase can improve classification of future data \parencite{tuckerGenerativeModelsFunctional2013}.

Recently, research has also been done on pairwise temporal alignment using neural networks. \textcite{nunezDeepLearningWarping2020a} proposed a convolutional neural network for the alignment of all pairs of functions in the data, which outputs estimated warping functions. However, the training data in this particular setting were constructed using DTW, the optimal warping functions from DWT were used in the loss function, and as the authors themselves expressed, their primary motivation was demonstrating reduced computational cost compared to DTW. In addition, the neural network was constructed to deal with pairwise alignment of all functional pairs, but not for joint alignment of multiple functions.

\textcite{ohLearningExploitInvariances2018} proposed an end-to-end classification model using a Sequence Transformer Network (STN), which transforms the input signals with parameters learned through a Sequence Transformer convolutional neural network, before proceeding with another neural network for classification. However, this STN focuses on linear transformation and does not implement more complicated time warping of the input signal.

In addition to unsupervised single-class learning, \textcite{weberDiffeomorphicTemporalAlignment2019} explored the multi-class case with semi-supervised learning. The implemented model is trained using a loss function that involves the empirical variance of the warped signals and uses a regularization term on the warping as a way to ensure identifiability.

Currently, more and more deep learning methods are developed to align multivariate functions using elastic phase-amplitude separation algorithm; see e.g., the works of \textcite{lohitTemporalTransformerNetworks2019a},  \textcite{nunezSrvfNetGenerativeNetwork2021} and \textcite{chenSrvfRegNetElasticFunction2021}, all of which use a probabilistic simplex transformation for the derivative of the warping functions by calculating the elementwise product of the neural network output and dividing it by the square of its norm. This transformation is, however, not one-to-one.

\subsection{Proposed approach}
In this paper, we propose a non-parametric approach for joint alignment of multivariate quasi-periodic functions. We deploy neural networks that aim to extract the optimal warping functions that achieve the smallest distance between the warped functions obtained from the observed functions by warping their domain. We use a special one-to-one activation function in the output layer, so that the output of the neural network satisfies the conditions for warping functions. In addition, instead of using the $\mathrm{L}^2$ metric to calculate the distance between functions directly, we use the Fisher-Rao metric and the square-root slope representation to calculate the distance between the warped functions. This metric is used in the loss function of the neural network. Furthermore, because we take advantage of the differential geometry of the space of warping functions, we can calculate the Karcher mean of the orbits to extract a functional template. 

We make use of convolutional layers with multiple channels to account for multivariate functions, and as a result, the output layer returns a single warping function per subject that can be used for warping all the dimensions of the observed function simultaneously, which is desirable as they are all measured on the same subject. Finally, we incorporate a multiscale warping model to handle quasi-periodic functions.

This paper is structured as follows: In Section~\ref{sec: joint phase amplitude}, we present and review the mathematical formalism behind joint alignment of functional data, which we proceed to extend in Section~\ref{sec: method} to the case of multivariate quasi-periodic functional data. In the same section, we present the architecture of the deep neural warping network employed in the algorithm. In Section~\ref{sec: simulation study}, we apply our method to simulated univariate and multivariate functions. In Section~\ref{sec: application}, we carry out the joint alignment on real ECG data, and finally in Section~\ref{sec: discussion and conclusion}, we discuss the results and limitations. An implementation of our method can be found on the first author's GitHub repository.

\section{Joint alignment of multivariate data}\label{sec: joint phase amplitude}
In this section, we present the mathematical background for joint alignment of a set of functions $\{f_i\}=\{f_i\}_{i=1}^{n} = \{f_i\in\mathcal{F}_I\mid i=1,\dots,n\}$, where $\mathcal{F}_I$ is the set of absolutely continuous functions defined on the interval $I$, which we without loss of generality take to be $I=[0,1]$. Further, we consider multivariate functional data $f_i=\left(f_{i1},\dots,f_{iJ}\right)\colon\, I\to\mathbb{R}^J$. To jointly align a set of functions, we need to warp the domain of the functions in a certain constrained way. The constraints for the mapping $\gamma\colon\, I\to I$ performing the domain warp are that $\gamma(0)=0$ and $\gamma(1)=1$ (\textit{boundary-preserving}), that $\gamma$ is invertible, and that both $\gamma$ and $\gamma^{-1}$ are differentiable (\textit{diffeomorphism}). We denote the set of such functions $\Gamma_I$. Note that the derivative of $\gamma$, denoted as $\dot{\gamma}$, is always positive.

With the definition of the two sets, $\mathcal{F}_I$ and $\Gamma_I$, we can describe the two main alignment problems: pairwise and multiple alignment \parencite[p.~85]{srivastavaFunctionalShapeData2016}. In the pairwise alignment problem, given functions $f_1,f_2\in\mathcal{F}_I$, we wish to find a warping function $\gamma\in\Gamma_I$, such that some energy term $E[f_1,f_2\circ\gamma]$, where the symbol ``$\circ$'' represents function composition, i.e., $(f\circ\gamma)(t)=f(\gamma(t))$, is minimized. In the case of multivariate functions, i.e., $J\geq2$, we calculate the function composition elementwise, namely, $f\circ\gamma=\left(f_1\circ\gamma,\dots,f_J\circ\gamma\right)$. The operations defined below are also elementwise, where relevant. We wish to find $\gamma^*$ as a solution to $\gamma^*=\argmin_{\gamma\in\Gamma_I}E[f_1,f_2\circ\gamma]$. The function $f_1$ is then said to be registered to $f_2\circ\gamma^*$ for any domain value $t\in I$. The multiple alignment problem is an extension of the pairwise alignment problem, where, given a set of functions $\{f_i\}$, we wish to find a set of warping functions $\{\gamma_i\}=\{\gamma_i\}_{i=1}^n=\{\gamma_i\in\Gamma_I\mid i=1,\dots,n\}$, such that $f_i\circ\gamma_i$ are said to be registered to each other over $i$. The $\{\gamma_i\}$ are called the \textit{phases} and $\{f_i\circ\gamma_i\}$ are representatives of their \textit{amplitude}. Finally, the pairwise solution of the multiple alignment, in which all pairs of functions are registered to each other, can be extended to a template-based registration. Here, we first consider a template function $\mu$, and then we align each of the functions $\{f_i\}$ to this template. These steps can be done iteratively to improve the overall alignment as well as the quality of the template \parencite[p.~271]{srivastavaFunctionalShapeData2016}. First, we calculate the average of the current versions of $\{f_i\}$ under a proper metric to construct the template $\mu$. Then we align the functions $\{f_i\}$ to this template by calculating the optimal warping functions $\{\gamma_i\}$, and update the functions $\{f_i\}$ by $f_i\gets f_i\circ\gamma_i$, and then we iterate these steps.

A natural way to calculate the average of $\{f_i\}$ for constructing the template would be to use the $\mathrm{L}^2$ norm, yielding the cross-sectional mean of $\{f_i\}$. However, as argued by \textcite[chap.~8.2]{srivastavaFunctionalShapeData2016}, the cross-sectional mean of $\{f_i\}$ is not a good representation of the template $\mu$, and in addition, also shown by \textcite[pp.~88--90]{srivastavaFunctionalShapeData2016}, the standard $\mathrm{L}^2$ norm is not appropriate as the distance measure due to the lack of isometry under warping, pinching effect, and inverse inconsistency. Instead, we use a specific Riemannian metric, called the Fisher-Rao metric \parencite[Definition 4.8., p.~105]{srivastavaFunctionalShapeData2016}, together with an alternative functional representation called the square-root slope function representation \parencite[SRSF,][Definition 4.2., p.~91]{srivastavaFunctionalShapeData2016} defined as
\begin{equation}\label{eq: SRSF}
    q(t)=\sign\left(\dot{f}(t)\right)\sqrt{\left|\dot{f}(t)\right|}.
\end{equation}
from which the original function can be recovered as
\begin{equation}\label{eq: original function}
    f(t)=f(0) + \int_0^tq(s)|q(s)|\diff s.
\end{equation}
Under this representation, the Fisher-Rao metric becomes the standard $\mathrm{L}^2$ metric \parencite[Lemma 4.7., p.~105]{srivastavaFunctionalShapeData2016}. Note also that the SRSF representation of a warped function $f\circ\gamma$ is \parencite[p.~91]{srivastavaFunctionalShapeData2016}
\begin{equation}\label{eq: warping in SRSF space}
    \tilde{q}(t)=(q\circ\gamma)(t)\sqrt{\dot{\gamma}(t)}\eqqcolon(q,\gamma)(t).
\end{equation}

\textcite[chap.~8]{srivastavaFunctionalShapeData2016} present an alternative estimator of $\mu$ which uses the notion of the amplitude of a function in the SRSF space. Given a function $f\in\mathcal{F}_I$ and its associated SRSF representation, $q$ in Equation~\eqref{eq: SRSF}, the amplitude of $f$ in the SRSF space is defined by the orbit, which is the set of all possible domain transformations according to the group action
\begin{equation*}
    [q]=\closure\left\{(q,\gamma)=(q\circ \gamma)\sqrt{\dot{\gamma}}\bigm|\gamma\in\Gamma_I\right\}.
\end{equation*}
With the definition of the amplitude, \textcite{srivastavaFunctionalShapeData2016} show that the Karcher mean of the set of orbits $\{[q_i]\}$, as described later in this section, is an estimator of the orbit $[\mu_q]$ and that a specific element of this orbit can be used as an estimator of $\mu_q$, which is the SRSF representation of $\mu$. The template-based alignment problem is then reduced to finding the set $\{\gamma_i\}$ that best aligns the functions $\{q_i\}$ to the template $\mu_q$, which can be formally written as
\begin{equation}\label{eq: template-based registration problem}
    \gamma_i=\underset{\gamma\in\Gamma_I}{\arginf}\|\mu_q-(q_i,\gamma)\|.
\end{equation}

In order to estimate $\mu_q$, we first have to calculate the Karcher mean of the set of orbits $\{[q_i]\}$ that is defined as \parencite[Definition 8.1., p.~274]{srivastavaFunctionalShapeData2016}
\begin{equation*}
[\mu_q]=\arginf_{[q]\in\mathcal{A}}\sum_{i=1}^n\inf_{\gamma\in\Gamma_I}(\|q-(q_i,\gamma)\|)^2, 
\end{equation*}
where $\mathcal{A}$ is a quotient space $\mathcal{F}_I/\tilde{\Gamma}_I$ and $\tilde{\Gamma}_I$ is a set of boundary preserving weakly increasing absolutely continuous functions $\gamma\colon\, I\to I$.

The Karcher mean of the amplitudes is again an orbit, and we have to find a particular element of this orbit, specifically its center with respect to the set $\{q_i\}$ \parencite[Definition 8.2., p.~275]{srivastavaFunctionalShapeData2016}. Such an element $\mu_q$ satisfies the property that the Karcher mean of the warping functions $\{\gamma_i\}$, which are the solutions to Equation~\eqref{eq: template-based registration problem}, is the identity warp, $\gamma_{\textrm{id}}(t)=t$.

\sloppy

The algorithm for finding a center of an orbit with respect to the set $\{q_i\}$ follows \textcite[Algorithm~33., p.~277]{srivastavaFunctionalShapeData2016} with some modifications. In the first step of the algorithm, we select an element $\tilde{\mu}_q$ of the orbit $[\mu_q]$, i.e., the cross-sectional mean of $\{q_i\}$, and we find $\{\tilde{\gamma}_i\}$ by solving ${\tilde{\gamma}_i=\arginf_{\gamma\in\Gamma_I}(\|\tilde{\mu}_q-(q_i,\gamma)\|)}$. As opposed to using dynamic programming for this problem as in \textcite{srivastavaFunctionalShapeData2016}, we propose to use a convolutional neural network. A second step of the algorithm is calculating the Karcher mean $\mu_{\tilde{\gamma}}$ of the phases $\{\tilde{\gamma}_i\}$ and finding the center of the orbit $[\mu_q]$ with respect to the set $\{q_i\}$ by
\begin{equation}\label{eq: center of the orbit}
    \mu_q=(\tilde{\mu}_q,\mu_{\tilde{\gamma}}^{-1}). 
\end{equation}
\textcite{srivastavaFunctionalShapeData2016} show that for each $q_i$,
\begin{equation}\label{eq: optimal warping}
    \gamma_i=\tilde{\gamma}_i\circ\mu_{\tilde{\gamma}}^{-1}
\end{equation}
minimizes ${\|\mu_q - (q_i,\gamma)\|}$, hence these $\{\gamma_i\}$ are a solution to the joint template-based registration problem represented by Equation~\eqref{eq: template-based registration problem}. The center, $\mu_q$, of the orbit is the estimator of the SRSF of the template function $\mu$.

\fussy

\paragraph{Karcher mean of warping functions.}\label{sec: Karcher mean of warping functions}
To define the Karcher mean of a set of warping functions $\gamma_i$ under the Fisher-Rao metric, we will use the differential geometry of $\Gamma_I$. Direct analysis on $\Gamma_I$ is not straightforward due to it being a non-linear manifold, thus we will work with the SRSF representation of $\gamma_i$. The SRSF representation of any $\gamma\in\Gamma_I$ has the form $\psi=\sqrt{\dot{\gamma}}$, which is equivalent to Equation~\eqref{eq: SRSF} because $\dot{\gamma} > 0$ for all domain values $t\in I$. An advantage of using this representation is that ${\|\psi\|^2=\int_0^1\psi^2(t)\diff t=\int_0^1\dot{\gamma}(t)\diff t=\gamma(1)-\gamma(0)=1}$, and thus $\psi$ lies in the positive orthant of the unit Hilbert sphere, $\mathbb{S}^\infty_+ =\left\{\psi\in\mathrm{L}^2\bigm|\|\psi\|=1, \psi > 0\right\}$. On $\mathbb{S}^\infty_+$, any point $\tilde{\psi}\in\mathbb{S}^\infty_+$ can be projected to the tangent space $T_\psi(\mathbb{S}^\infty_+)=\left\{v\in\mathrm{L}^2\bigm|\int_0^1\psi(t) v(t)\diff t=0\right\}$ at $\psi$ by the inverse exponential map \parencite[p.~83]{srivastavaFunctionalShapeData2016} by
\begin{equation}\label{eq: inverse exponential map}
    \left(\exp_\psi^{-1}\tilde{\psi}\right)(t)=\frac{\theta}{\sin(\theta)}\left(\psi(t)-\tilde{\psi}(t)\cos(\theta)\right), \quad t\in[0,1],
\end{equation}
with $\theta=\cos^{-1}\left(\int_0^1\psi(t)\tilde{\psi}(t)\diff t\right)$.
Similarly, points in the tangent space can be projected back to the unit sphere $\mathbb{S}^\infty_+$ at the point $\psi$ by the exponential map \parencite[p.~83]{srivastavaFunctionalShapeData2016}
\begin{equation}\label{eq: exponential map}
    \left(\exp_\psi v\right)(t)=\cos\left(\|v\|\right)\psi(t)+\sin\left(\|v\|\right)\frac{v(t)}{\|v\|},\quad t\in[0,1].
\end{equation}
The algorithm for finding the Karcher mean of warping functions under the Fisher-Rao metric follows \textcite[Algorithm~24, p.~238]{srivastavaFunctionalShapeData2016} and is formally described in Algorithm~\ref{alg: karcher mean of warping functions}, with an initial estimate of the Karcher mean of the warping functions being the normalized cross-sectional mean of their SRSFs. This procedure is based on a fixed-point algorithm, see e.g., \textcite[chap.~5]{bhattacharyaNonparametricInferenceManifolds2012}. 
\begin{algorithm}
\DontPrintSemicolon
\caption{Karcher mean of warping functions}\label{alg: karcher mean of warping functions}
\KwInput{Warping functions $\{\gamma_i\}_{i=1}^n$, stopping criterion $c$, maximum number of iterations $E$, step size $\epsilon$}
\KwOutput{Karcher mean $\mu_\gamma$}
\vspace{-0.5em}\nonl\hrulefill\\
$\psi_i\gets\SRSF(\gamma_i)$\KwComment*[r]{Calculate SRSF representation of $\gamma_i$ using \eqref{eq: SRSF}}
$\mu_{\psi}\gets\frac{1}{n}\sum_{i=1}^n\psi_i$\KwComment*[r]{Initialize mean of SRSFs}
$\mu_{\psi}\gets\mu_{\psi}/\|\mu_{\psi}\|$\KwComment*[r]{Normalize the mean of SRSFs}
$\mu_v\gets 0$\KwComment*[r]{Initialize mean in the tangent space}
$e\gets0$\KwComment*[r]{Initialize the index of iteration}
\While{$(\|\mu_v\|\geq c$ \KwOr $e=0)$ \KwAnd $e<E$}{
$e\gets e+1$\KwComment*[r]{Increase the index of iteration}
$\mu_\psi\gets\exp_{\mu_\psi}\epsilon\mu_v$\KwComment*[r]{Update mean of SRSFs using \eqref{eq: exponential map}}
$v_i\gets\exp_{\mu_\psi}^{-1}\psi_i$\KwComment*[r]{Project $\psi_i$ to tangent space at $\mu_\psi$ using \eqref{eq: inverse exponential map}}
$\mu_v\gets\frac{1}{n}\sum_{i=1}^nv_i$\KwComment*[r]{Calculate mean of projections $\{v_i\}$}
}
$\mu_\gamma\gets\ToWarp(\mu_\psi)$\KwComment*[r]{Calculate mean warping function using \eqref{eq: original function}}
\end{algorithm}

\subsection{Joint alignment of multivariate quasi-periodic data}\label{sec: method}
In this section, we describe the extension of the joint alignment algorithm to multivariate quasi-periodic functions. We define a periodic extension of a function $f\colon[0,\tau]\to\mathbb{R}$ as $\ext^K_{\mathrm{L}^2}f\colon\,[0,K\tau]\to\mathbb{R}$, $K\in\mathbb{N}$ with
\begin{equation}\label{eq: periodic extension - function}
    \left(\ext^K_{\mathrm{L}^2}f\right)(t+k\tau)\coloneqq f(t)\quad\textrm{and}\quad\left(\ext^K_{\mathrm{L}^2}f\right)(K\tau)\coloneqq f(\tau),
\end{equation}
for $t\in[0,\tau)$ and $k=0,\dots,K-1$. We also define a periodic extension of a warping function $\gamma\colon\,[0,\tau]\to[0,\tau]$ on $[0,K\tau]$ by
\begin{equation}\label{eq: periodic extension - gamma}
    \left(\ext^K_{\Gamma}\gamma\right)(t+k\tau)\coloneqq\gamma(t)+k\tau\quad\textrm{and}\quad\left(\ext^K_{\Gamma}\gamma\right)(K\tau)\coloneqq K\tau,
\end{equation}
for $t\in[0,\tau)$ and $k=0,\dots,K-1$. As a dual to the periodic extension, we define a split of a function ${f\colon\,[0,K\tau]\to\mathbb{R}}$ in its domain into $K$ functions with $t\in[0,\tau]$ by
\begin{equation}
    \begin{aligned}\label{eq: split}
        \left(\spl^Kf\right)(t)
        &=\left(f(t),\dots,f(t+(k-1)\tau),\dots,f(t+(K-1)\tau)\right)\\
        &\coloneqq\left(f_1(t),\dots,f_k(t),\dots,f_K(t)\right)),
    \end{aligned}
\end{equation}
for $k=1,\dots,K$.
After performing the extension or the split of a warping function, we need to linearly transform the image of the resulting warping functions to the same interval as the domain in order to keep the boundary-preserving property, and furthermore, whenever the exponential and inverse exponential maps, Equations~\eqref{eq: exponential map} and \eqref{eq: inverse exponential map}, respectively, need to be used, the domain and image of the warping functions need to be transformed to the interval $[0,1]$.

\sloppy

\begin{figure}
    \centering
    \begin{subfigure}[b]{0.33\textwidth}
        \centering
        \includegraphics[width=\textwidth]{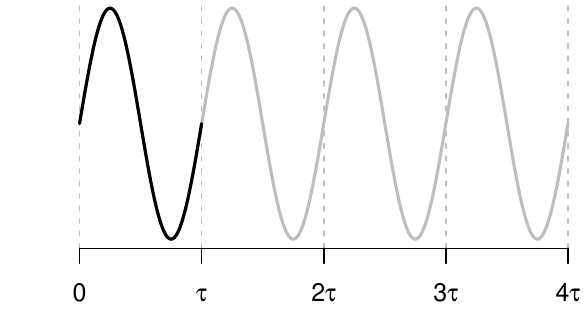}
        \caption{Common template}
        \label{fig: common template}
    \end{subfigure}%
    \begin{subfigure}[b]{0.33\textwidth}
        \centering
        \includegraphics[width=\textwidth]{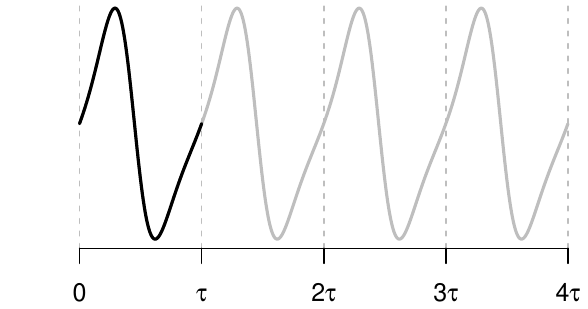}
        \caption{Subject-specific template}
        \label{fig: subject-specific template}
    \end{subfigure}%
    \begin{subfigure}[b]{0.33\textwidth}
        \centering
        \includegraphics[width=\textwidth]{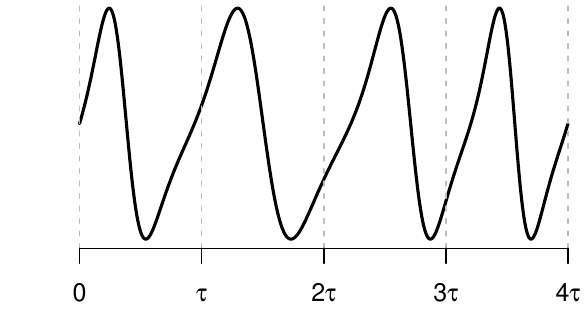}
        \caption{Observed function}
        \label{fig: observed function}
    \end{subfigure}
    \begin{subfigure}[b]{0.33\textwidth}
        \centering
        \includegraphics[width=\textwidth]{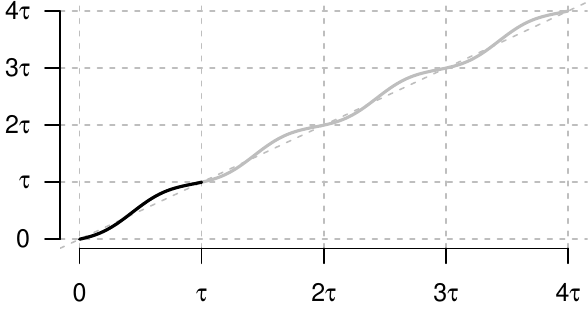}
        \caption{Local warping}
        \label{fig: subject-specific local warping}
    \end{subfigure}%
    \begin{subfigure}[b]{0.33\textwidth}
        \centering
        \includegraphics[width=\textwidth]{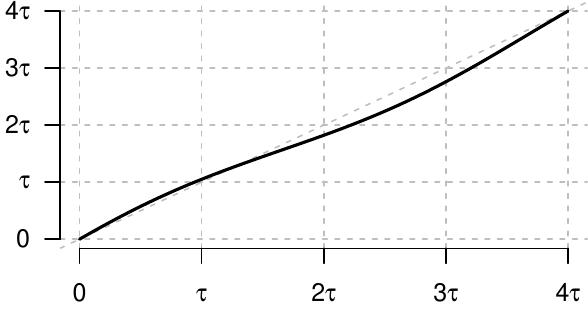}
        \caption{Global warping}
        \label{fig: subject-specific global warping}
    \end{subfigure}%
    \begin{subfigure}[b]{0.33\textwidth}
        \centering
        \includegraphics[width=\textwidth]{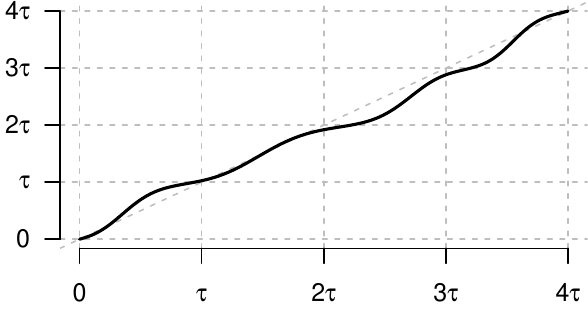}
        \caption{Total warping}
        \label{fig: subject-specific total warping}
    \end{subfigure}
    \caption{Multiscale warping model. The common template is represented by the black line on $[0,\tau]$ in \ref{fig: common template}, where the gray line represents the periodic extension of this template on $[0,4\tau]$. The subject-specific template is shown in \ref{fig: subject-specific template}, which is a result of warping the common template with the subject-specific local warping function, as represented in \ref{fig: subject-specific local warping}. \ref{fig: observed function} shows the observed function that was obtained from the periodic extension of the subject-specific template warped with the subject-specific global warping function, as represented in \ref{fig: subject-specific global warping}. Finally, \ref{fig: subject-specific total warping} shows the composition of the subject-specific local and the subject-specific global warping functions.}
    \label{fig: multiscale warping model}
\end{figure}

The observed multivariate quasi-periodic functions $f_i\colon\,[0,K\tau]\to\mathbb{R}^J$, where ${f_i=\left(f_{i1},\dots,f_{iJ}\right)}$ with their SRSF representations $q_i=\left(q_{i1},\dots,q_{iJ}\right)$, are assumed to be generated from a multiscale warping model by
\begin{equation}
\begin{aligned}\label{eq: multiscale warping model}
    f_{ij}(t)&=\left(\ext^K_{\mathrm{L}^2}\mu_j\circ\ext^K_{\Gamma}\gamma_i^l\circ\gamma_i^g\right)\left(t\right)\\
    &=\left(\ext^K_{\mathrm{L}^2}\mu_{ij}\circ\gamma_i^g\right)\left(t\right)\\
    &=\left(\ext^K_{\mathrm{L}^2}\mu_j\circ\gamma_i^t\right)\left(t\right),
\end{aligned}
\end{equation}
where $\gamma_i^l\colon\,[0,\tau]\to[0,\tau]$ is the \textit{local} warping function and $\ext^K_{\Gamma}\gamma_i^l\colon\,[0,K\tau]\to[0,K\tau]$ its periodic extension, $\gamma_i^{g}\colon\,[0,K\tau]\to[0,K\tau]$ is the \textit{global} warping function, and $\gamma_i^{t}\colon\,[0,K\tau]\to[0,K\tau]$ is the \textit{total} warping function defined as the composition ${\ext^K_{\Gamma}\gamma_i^l\circ\gamma_i^{g}}$, see Figure \ref{fig: multiscale warping model}. We think of $\mu=\left(\mu_1,\dots,\mu_J\right)$ as the common template and ${\mu_i=\mu\circ\gamma_i^l=\left(\mu_1\circ\gamma_i^l,\dots,\mu_J\circ\gamma_i^l\right)=\left(\mu_{i1},\dots,\mu_{iJ}\right)}$ as the subject-specific template on $[0,\tau]$.

\fussy

On the other hand, given functions $\gamma_i^l$ and $\gamma_i^g$, or $\gamma_i^t$, we can extract the periodic extension $\ext^K_{\mathrm{L}^2}\mu_j$ of the common template function $\mu_j$ from an observed function $f_{ij}$ by
\begin{equation}
\begin{aligned}\label{eq: multiscale warping model - inverse}
    \left(\ext^K_{\mathrm{L}^2}\mu_j\right)(t)&=\left(f_{ij}\circ\left(\gamma_i^t\right)^{-1}\right)(t)\\
    &=\left(f_{ij}\circ\left(\gamma_i^g\right)^{-1}\circ\left(\ext^K_{\Gamma}\gamma_i^{l}\right)^{-1}\right)(t)\\
    &=\left(\ext^K_{\mathrm{L}^2}\mu_{ij}\circ\left(\ext^K_{\Gamma}\gamma_i^{l}\right)^{-1}\right)(t).
\end{aligned}
\end{equation}

\subsection{Algorithm for joint alignment of multivariate quasi-periodic data using deep learning}\label{sec: DeepJAM algorithm}
\begin{figure}
    \centering
    \includegraphics{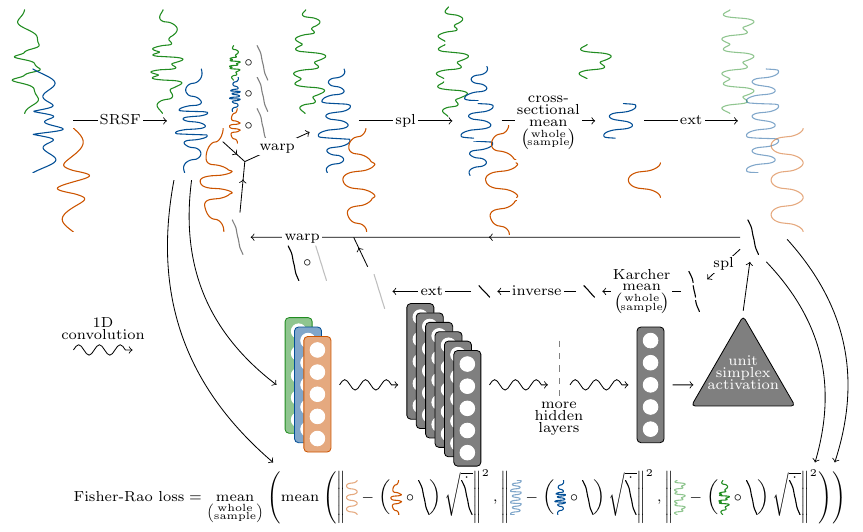}
    \caption{Diagram of the algorithm for joint alignment of multivariate quasi-periodic functions. Note that both the cross-sectional and Karcher mean are taken over the whole sample. Furthermore, the domain and image of the warping functions that are involved in the Karcher mean are first scaled to the interval $[0,1]$.}
    \label{fig: algorithm}
\end{figure}

The algorithm for adaptive template-based groupwise registration follows \textcite[Algorithm~33., p.~277]{srivastavaFunctionalShapeData2016} with some modifications, and is represented in Algorithm~\ref{alg: DeepJAM}, as well as in Figure~\ref{fig: algorithm}. The most important modifications are replacing dynamic programming \parencite[Algorithm~58., p.~437]{srivastavaFunctionalShapeData2016} with a convolutional neural network, and extending the algorithm to handle quasi-periodic multivariate functions.
\begin{algorithm}
\DontPrintSemicolon
\caption{Joint alignment of multivariate quasi-periodic functional data using deep learning: DeepJAM}\label{alg: DeepJAM}
\algorithmfootnote{$\NN(x=\{q_i\},y=\mu_{\tilde{q}}^K)$ and $\NN(x=q_i)$ represent a neural network with input $\{q_i\}$ and $q_i$, respectively, and outcome $\mu_{\tilde{q}}^K$.}
\KwInput{Observed functions $\{f_i\}_{i=1}^n$ on the interval $I=[0,1]$, number of periods~$K$, number of iterations $E$, initialized neural network $\NN$}
\KwOutput{Warping functions $\{\gamma_i\}_{i=1}^n$, SRSF $\mu_q$ of the common template}
\vspace{-0.5em}\nonl\hrulefill\\
$\gamma_i\gets\gamma_{\textrm{id}}$\KwComment*[r]{Initialize $\gamma_i$ to identity warp}
$q_i\gets\SRSF(f_i)$\KwComment*[r]{Calculate SRSF representation of $f_i$ using \eqref{eq: SRSF}}
\For{$e$ \KwFrom $1$ \KwTo $E$}{
$\tilde{q}_i\gets(q_i,\gamma_i)$\KwComment*[r]{Warp in the SRSF space using \eqref{eq: warping in SRSF space}}
$\left(\tilde{q}_{i1},\dots,\tilde{q}_{iK}\right)\gets\spl^K\tilde{q}_i$\KwComment*[r]{Split the SRSFs using \eqref{eq: split}}
$\mu_{\tilde{q}}\gets\frac{1}{nK}\sum_{i=1}^n\sum_{k=1}^K\tilde{q}_{ik}$\KwComment*[r]{Calculate mean of $\{\tilde{q}_{ik}\}$}
$\mu_{\tilde{q}}^K\gets\ext^K_{\mathrm{L}^2}\mu_{\tilde{q}}$\KwComment*[r]{Calculate extension of $\mu_{\tilde{q}}$ using \eqref{eq: periodic extension - function}}
train $\NN(x=\{q_i\},y=\mu_{\tilde{q}}^K)$ for one epoch\\
$\tilde{\gamma}_i\gets\NN(x=q_i)$\KwComment*[r]{Calculate neural network prediction}
$\left(\tilde{\gamma}_{i1},\dots,\tilde{\gamma}_{iK}\right)\gets\spl^K\tilde{\gamma}_i$\KwComment*[r]{Split warping functions using \eqref{eq: split}}
$\tilde{\gamma}_{ik}^\prime\gets\Scale(\tilde{\gamma}_{ik})$\KwComment*[r]{Scale domain and image of $\tilde{\gamma}_{ik}$ to $[0,1]$}
$\mu_{\tilde{\gamma}}\gets\KarcherMean(\{\tilde{\gamma}_{ik}^\prime\})$\\\KwComment*[r]{Find the Karcher mean of $\{\tilde{\gamma}_{ik}^\prime\}$ as in Algorithm~\ref{alg: karcher mean of warping functions}}
$\gamma^*\gets\ext^K_{\Gamma}\mu_{\tilde{\gamma}}^{-1}$\KwComment*[r]{Calculate extension of $\mu_{\tilde{\gamma}}^{-1}$ using \eqref{eq: periodic extension - gamma}}
$\gamma^\prime\gets\Scale\left(\gamma^*\right)$\KwComment*[r]{Scale domain and image of $\gamma^*$ to $[0,1]$}
$\gamma_i\gets\tilde{\gamma}_i\circ\gamma^\prime$\KwComment*[r]{Calculate warping functions using \eqref{eq: optimal warping}}
}
$\gamma^\prime\gets\Scale\left(\mu_{\tilde{\gamma}}^{-1}\right)$\KwComment*[r]{Scale domain and image of $\mu_{\tilde{\gamma}}^{-1}$ to [0,1/K]}
$\mu_q\gets(\mu_{\tilde{q}},\gamma^\prime)$\KwComment*[r]{Calculate SRSF of common template using \eqref{eq: center of the orbit}}
\end{algorithm}

\paragraph{Subject-specific template.}

\sloppy

The warping functions $\{\gamma_i\}$ obtained by Algorithm~\ref{alg: DeepJAM} correspond to the inverse of the total warping functions as described in Equations~\eqref{eq: multiscale warping model} and \eqref{eq: multiscale warping model - inverse}. In order to obtain subject-specific templates, we need to decompose this total warping function into the global and local warping functions, and we propose that this can be done in the following way. Consider the subject $i$ and the corresponding warping function $\gamma_i$ that aligns the observed function $f_i$ to the periodic extension of the common template, $\ext^K_{\mathrm{L}^2}\mu$. We split the domain of the warping function $\gamma_i$ into $K$ functions using Equation~\eqref{eq: split}. This yields $\left(\gamma_{i1},\dots,\gamma_{iK}\right)$ which warp the split of $q_i$, $\left(q_{i1},\dots,q_{iK}\right)$, to an element of an orbit $[\mu_{q_i}]$. We define the SRSF of the subject-specific template as the center of the orbit $[\mu_{q_i}]$ with respect to the set $\{q_{ik}\}_{k=1}^K$. We find this by calculating the Karcher mean $\mu_{\gamma_i}$ of $\{\gamma_{ik}\}_{k=1}^K$ by first scaling their domain and image to the interval $[0,1]$, and then performing Algorithm~\ref{alg: karcher mean of warping functions}. We can represent the warping function $\gamma_i$ as
\begin{equation}\label{eq: decomposition}
    \gamma_i=\gamma_i\circ\gamma_{\textrm{id}}=\gamma_i\circ\left(\ext^K_{\Gamma}\mu_{\gamma_i}\right)^{-1}\circ\ext^K_{\Gamma}\mu_{\gamma_{i}},
\end{equation}
where $\ext^K_{\Gamma}\mu_{\gamma_{i}}$ is a periodic extension of $\mu_{\gamma_{i}}$, with the domain and image scaled to the interval $[0,1]$. Furthermore, we have the following identities 
\begin{equation}\label{eq: identities}
    \gamma_i = \left(\gamma_i^t\right)^{-1},\quad
    \gamma_i\circ\left(\ext^K_{\Gamma}\mu_{\gamma_i}\right)^{-1} = \left(\gamma_i^g\right)^{-1},\quad
    \mu_{\gamma_{i}} = \left(\gamma_i^l\right)^{-1},
\end{equation}
where $\gamma_i^t$, $\gamma_i^g$ and $\gamma_i^l$ are the total, global and local warping functions, respectively, as described in Equation~\eqref{eq: multiscale warping model}. 

The SRSF of the common template, $\mu_q$, can be obtained from Algorithm~\ref{alg: DeepJAM}, and the common template $\mu$ can be obtained from its SRSF using Equation~\eqref{eq: original function} up to a constant $c\in\mathbb{R}$, $\mu(t)=c+\int_0^t\mu_q(s)|\mu_q(s)|\diff s$. The subject-specific template can be extracted using identities in Equations~\eqref{eq: multiscale warping model} and \eqref{eq: identities} by
\begin{equation}\label{eq: subject-specific template}
    \mu_{i}=\mu\circ\mu_{\gamma_i}^{l}=\mu\circ\mu_{\gamma_i}^{-1}.
\end{equation}

\fussy

\paragraph{Quasi-periodic functions with amplitude variability.}
In addition to the multiscale time warping model, the amplitude of the subject-specific template can be further susceptible to amplitude variability, which means that the amplitudes in the different periods might not be the same. Furthermore, amplitudes measured on different subjects do not have to be the same either, thus there is no longer the same relationship between $\mu_j$ and $\mu_{ij}$ as described in Equation~\eqref{eq: multiscale warping model - inverse}, and the subject-specific template cannot be obtained by Equation~\eqref{eq: subject-specific template}.

\sloppy

To obtain the SRSF of the subject-specific template in the presence of amplitude variability, we propose calculating the cross-sectional mean of the aligned functional data
\begin{equation*}
    \mu_{q_i}(t)=\frac{1}{K}\sum_{k=1}^K(q_{ik},\gamma_{ik}^\prime)(t),
\end{equation*}
where $t\in[0,1]$, $\{\left(q_{i1},\dots,q_{iK}\right)\}$ is the split of $q_i$ using Equation~\eqref{eq: split}, and $\{\left(\gamma_{i1}^\prime,\dots,\gamma_{iK}^\prime\right)\}$ is the split of $\gamma_i\circ\left(\ext^K_{\Gamma}\mu_{\gamma_i}\right)^{-1}$ using Equation~\eqref{eq: split}. Finally, the subject-specific template can be obtained up to a constant from $\mu_{q_i}$ using Equation~\eqref{eq: original function}.

\fussy

\subsection{Deep learning architecture}\label{sec: neural network architecture}
The general architecture of a warping neural network can be quite flexible. The most important part is that it produces an output that has similar dimension of the discretized input functions and can be transformed into a warping function. In our case, this is achieved by using convolutional layers with a padding such that the layer output has the same length as the layer input. The outputs of the neural network are the warping functions, which do not need to have exactly the same dimension as the input layer. In this case, we can use interpolation to scale the output up or down. Furthermore, to save computation time and because of the specific loss function described below, instead of inputting the original functions, we use their SRSF representation as inputs directly. We implemented the neural network using Keras \parencite{chollet2015keras} and TensorFlow \parencite{tensorflow2015-whitepaper} in the R packages \texttt{keras} \parencite{keras} and \texttt{tensorflow} \parencite{tensorflow}.

\sloppy

The output of the neural network uses a special activation function based on the unit simplex transformation, so that the resulting warping functions are boundary-preserving increasing functions on the interval $[0,1]$. For a given neural network output $y\in\mathbb{R}^P$ before any activation is applied, we calculate new coordinates $z\in\mathbb{R}^P$ by
\begin{equation}\label{eq: z coordinates}
z_p=\expit\left(y_p - \log(P-p+1)\right),\,\textrm{for}\,\,1\leq p\leq P,
\end{equation}
where $\expit(x)=1/(1+\exp(-x))$, ensuring that the new coordinates are all between 0 and 1. The vector $z$ is then used to determine a vector $x\in\mathbb{R}^{P+1}$ by
\begin{equation*}\label{eq: unit simplex}
x_p=
\begin{cases}
z_p,&p=1,\\
\left(1-\sum_{p^\prime=1}^{p-1}x_{p^\prime}\right)z_p,&\,1< p\leq P,\\
1-\sum_{p^\prime=1}^{P}x_{p^\prime},&\,p=P+1,
\end{cases}
\end{equation*}
with the property that all the coordinates of $x$ are greater than 0, and that ${\sum_{p=1}^{P+1}x_p=1}$. Looking back at Equation~\eqref{eq: z coordinates}, the offset $-\log(P-p+1)$ is added so that the zero vector $y$ is mapped to the simplex $x=(1/(P+1),\dots,1/(P+1))$. Finally, we transform this vector $x$ to a discretized warping function $\gamma\in\mathbb{R}^{P+2}$ by
\begin{equation*}\label{eq: simplex to warping}
\gamma_p=
\begin{cases}
0,&p=1,\\
\sum_{p^\prime=1}^{p-1}x_{p^\prime}&1<p\leq P+2.
\end{cases}
\end{equation*}
In order to obtain a vector in $\mathbb{R}^{P}$, we propose linearly interpolating the values of $\gamma$, so that $\gamma\in\mathbb{R}^{P}$.

\fussy

The loss function of our warping neural network is defined as follows. Let $f_i$, where $f_i(t)=\left(f_{i1}(t),\dots,f_{iJ}(t)\right)\in\mathbb{R}^J$, be the observed multivariate functions and $q_{ij}$ the SRSF representations of $f_{ij}$. Let $\mu_{q_j}$ be the SRSF representation of the template $\mu_{j}$, $j=1,\dots,J$, and let $\gamma_i$ be the outputs of the neural network. The Fisher-Rao loss function is then defined as
\begin{equation*}
    \mathcal{L}=\frac{1}{n}\sum_{i=1}^n\frac{1}{J}\sum_{j=1}^J\|\mu_{q_j}-(q_{ij},\gamma_i)\|^2.
\end{equation*}
The univariate function case is easily obtainable by setting $J=1$.

We used the Adam optimizer \parencite{kingmaAdamMethodStochastic2015}. The architectural hyperparameters such as the number of layers, number of filters, kernel size, and learning rate were chosen with the Bayesian optimization algorithm \parencite{snoekPracticalBayesianOptimization2012} using the expected improvement acquisition function. Finally, the hyperbolic tangent was used as an activation function of the hidden layers.

\section{Simulation study}\label{sec: simulation study}
As a proof-of-concept of our neural network approach, we deploy two simulation studies: one for univariate quasi-periodic functions and one for multivariate quasi-periodic functions. In the case of univariate functions, we did not vary the amplitudes, hence the functions can be perfectly aligned, i.e., the distance between the true template and the aligned functions converges to zero with increasing sample size and complexity of the neural network. In the case of multivariate functions we varied both the inter-subject and intra-subject amplitudes, thus the peaks and valleys of the functions can be temporally aligned to the common template, but the vertical distance between them does not converge to zero.

In both cases, we generated $N=14,000$ functions and used 8,000 of them as the training data, $2,000$ for hyperparameter tuning, $2,000$ as the validation data, and $2,000$ as the test data. 

To evaluate the performance of our method, we use the decrease in the cumulative cross-sectional variance of the observed and aligned data. The cumulative cross-sectional variance is a measure of the average distance of the functions from the mean, defined as follows. Let $\left\{f_i(t),t\in[0,1]\right\}_{i=1}^n$ be a functional dataset, then the cumulative cross-sectional variance is
\begin{equation*}\label{eq: CCSV mean}
    \widehat{\Var}\left(\{f_i\}\right)=\frac{1}{n-1}\int_0^1\sum_{i=1}^n\left(f_i(t)-\mu(t)\right)^2\diff t.
\end{equation*}
However, since we do not know the true common template $\mu$ for the real application data, we will use the cross-sectional mean in the calculation instead
\begin{equation*}\label{eq: CCSV}
    \overline{\Var}\left(\{f_i\}\right)=\frac{1}{n-1}\int_0^1\sum_{i=1}^n\left(f_i(t)-\frac{1}{n}\sum_{i=1}^n f_i(t)\right)^2\diff t.
\end{equation*}
A decrease in the cumulative cross-sectional variance is desirable, because the cumulative cross-sectional variance of the observed data contains both the amplitude and phase variability, whereas the cumulative cross-sectional variance of the aligned data represents only the variability of the amplitude. Furthermore, we calculate the square of the $\mathrm{L}^2$ norm of the distance between the cross-sectional mean and the true common template by
\begin{equation*}\label{eq: distance mean}
    \left\|\frac{1}{n}\sum_{i=1}^n f_i-\mu\right\|^2=\int_0^1\left(\frac{1}{n}\sum_{i=1}^n f_i(t)-\mu(t)\right)^2\diff t.
\end{equation*}

\subsection{Scenario 1, univariate functions}\label{sec: univariate functions}
In the first study, we simulated the common template to be the sine wave on the interval $[0,2\pi]$, $\mu_f=\sin(t)$, $t\in[0,2\pi]$, sampled at 65 equidistant points on each of the $K=3$ periods, adding up to $P=193$ equidistant points. Using the described notation, the periodic extension of the template is a sine wave on the interval $[0,6\pi]$, $\ext^K_{\mathrm{L}^2}\mu=\sin(t)$, $t\in[0,6\pi]$. The domain of the functions were then scaled to the interval $[0,1]$. We simulated $N=14,000$ global and $N=14,000$ local warping functions denoted $\gamma_i^g$ and $\gamma_i^l$ using the \texttt{fdasrvf} package \parencite{fdasrvf}, respectively. The warping functions using this package are generated in a way, so that their Karcher mean is the identity function. However, to ensure the identifiability of the subject-specific template, we need to transform the generated warping functions using Equation~\eqref{eq: decomposition} and the identities in Equation~\eqref{eq: identities}. The observed functions are then generated as $f_i(t)=\left(\ext^K_{\mathrm{L}^2}\mu\circ(\gamma_i^l)^K\circ\gamma_i^g\right)(t)$.

The optimized architecture consists of an input layer with 193 nodes and one channel, followed by 17 convolutional layers with a kernel of size 64. The hidden layers have 25 filters, while the output layer has only one filter. The learning rate was chosen to be $3.66\cdot10^{-7}$.

\begin{figure}
    \centering
    \begin{subfigure}{0.5\textwidth}
        \newcounter{figurescounter}
        \stepcounter{figurescounter}
        \centering
        \begin{subfigure}[b]{0.5\textwidth}
            \renewcommand\thesubfigure{\alph{subfigure}\thefigurescounter}
            \centering
            \includegraphics[width=\textwidth]{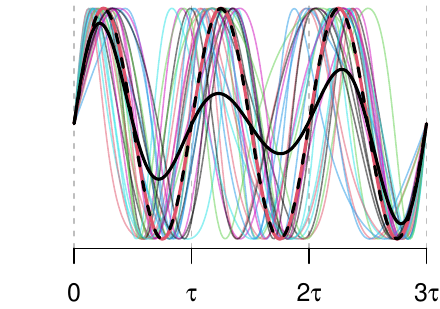}
            \caption{Observed}
            \label{fig: SIM one train observed}
        \end{subfigure}%
        \begin{subfigure}[b]{0.5\textwidth}
            \addtocounter{subfigure}{-1}
            \stepcounter{figurescounter}
            \renewcommand\thesubfigure{\alph{subfigure}\thefigurescounter}
            \centering
            \includegraphics[width=\textwidth]{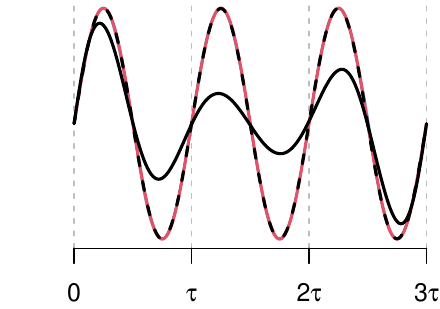}
            \caption{Aligned}
            \label{fig: SIM one train aligned}
        \end{subfigure}
        \begin{subfigure}[b]{0.5\textwidth}
            \addtocounter{subfigure}{-1}
            \stepcounter{figurescounter}
            \renewcommand\thesubfigure{\alph{subfigure}\thefigurescounter}
            \centering
            \includegraphics[width=\textwidth]{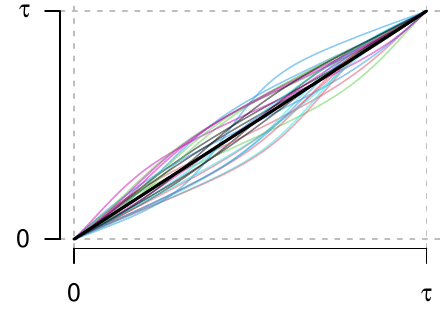}
            \caption{Local warping}
            \label{fig: SIM one train local}
        \end{subfigure}%
        \begin{subfigure}[b]{0.5\textwidth}
            \addtocounter{subfigure}{-1}
            \stepcounter{figurescounter}
            \renewcommand\thesubfigure{\alph{subfigure}\thefigurescounter}
            \centering
            \includegraphics[width=\textwidth]{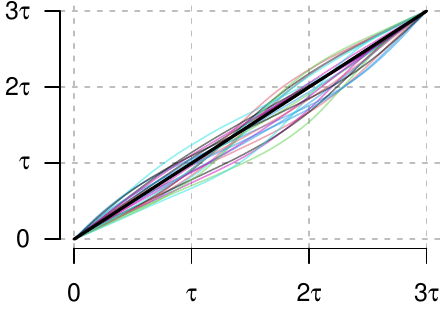}
            \caption{Global warping}
            \addtocounter{subfigure}{-1}
            \label{fig: SIM one train global}
        \end{subfigure}
        \caption{Training data}
        \label{fig: SIM one train}
    \end{subfigure}%
    \begin{subfigure}{0.5\textwidth}
        \setcounter{figurescounter}{0}
        \begin{subfigure}[b]{0.5\textwidth}
            \stepcounter{figurescounter}
            \renewcommand\thesubfigure{\alph{subfigure}\thefigurescounter}
            \centering
            \includegraphics[width=\textwidth]{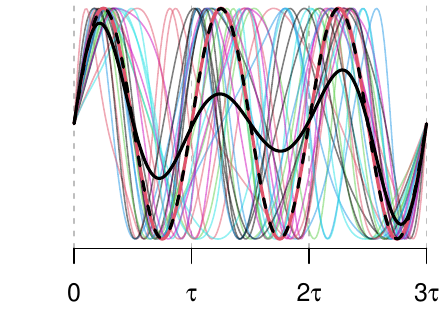}
            \caption{Observed}
            \label{fig: SIM one test observed}
        \end{subfigure}%
        \begin{subfigure}[b]{0.5\textwidth}
            \addtocounter{subfigure}{-1}
            \stepcounter{figurescounter}
            \renewcommand\thesubfigure{\alph{subfigure}\thefigurescounter}
            \centering
            \includegraphics[width=\textwidth]{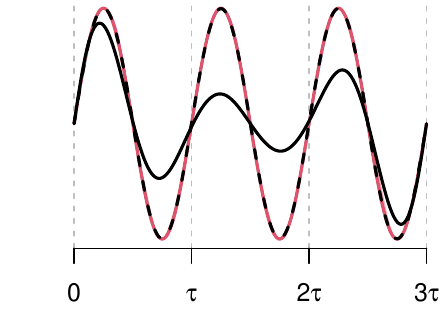}
            \caption{Aligned}
            \label{fig: SIM one test aligned}
        \end{subfigure}
        \begin{subfigure}[b]{0.5\textwidth}
            \addtocounter{subfigure}{-1}
            \stepcounter{figurescounter}
            \renewcommand\thesubfigure{\alph{subfigure}\thefigurescounter}
            \centering
            \includegraphics[width=\textwidth]{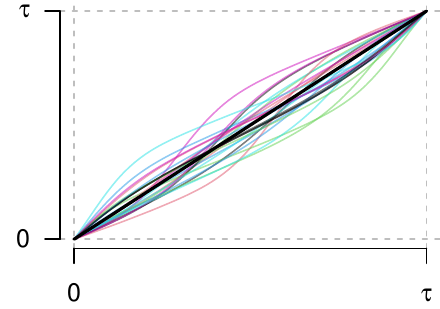}
            \caption{Local warping}
            \label{fig: SIM one test local}
        \end{subfigure}%
        \begin{subfigure}[b]{0.5\textwidth}
            \addtocounter{subfigure}{-1}
            \stepcounter{figurescounter}
            \renewcommand\thesubfigure{\alph{subfigure}\thefigurescounter}
            \centering
            \includegraphics[width=\textwidth]{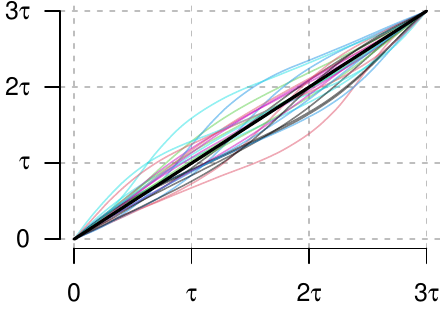}
            \caption{Global warping}
            \label{fig: SIM one test global}
        \end{subfigure}
        \addtocounter{subfigure}{-1}%
        \caption{Test data}
        \label{fig: SIM one test}
    \end{subfigure}
    \caption{Result of the DeepJAM algorithm. The first row shows the following. Figures~\ref{fig: SIM one train observed} and \ref{fig: SIM one test observed} show random 25 observed functions from the training and test data, respectively, while Figures~\ref{fig: SIM one train aligned} and \ref{fig: SIM one test aligned} show the same functions, but aligned using the DeepJAM neural network, respectively for the training and test data. The thick red lines are the true common template extended to three periods, and the thick black full/dashed lines represent the cross-sectional mean of the full observed/aligned training and test data, separately. The second row shows the estimated local and global warping of these random 25 functions from the training and test data, where the thick black lines are the identity lines.}
    \label{fig: SIM one}
\end{figure}

Figure \ref{fig: SIM one} shows the observed and aligned functions, as well as the corresponding estimated local and global warping functions. We can see that we achieved almost perfect alignment of both the training and test data. Furthermore, we can see that the cross-sectional mean of the observed data still preserves the modality in the data, but the size of the amplitudes is distorted. The cumulative cross-sectional variance of the simulated multivariate data can be found in Table~\ref{tab: SIM_one cumulative cross-sectional variance}.
\begin{table}
    \centering
    \begin{tabular}{rrrrrr}
  \toprule
\multicolumn{3}{c}{Functions}&\multicolumn{3}{c}{Mean}\\
\multicolumn{1}{c}{Observed} & \multicolumn{1}{c}{Aligned} & \multicolumn{1}{c}{$\%\downarrow$} & \multicolumn{1}{c}{Observed} & \multicolumn{1}{c}{Aligned} & \multicolumn{1}{c}{$\%\downarrow$} \\ 
  \midrule
$0.475$ & $0.001$ & $99.81$ & $0.148$ & $1.89 \cdot 10^{-5}$ & $99.99$ \\ 
   \bottomrule
\end{tabular}

    \caption{Cumulative cross-sectional variance of the observed and aligned simulated univariate data, where the columns ``$\%\downarrow$'' show the reduction of the cumulative cross-sectional variance in $\%$. In addition, we calculate the square of the $\mathrm{L}^2$ distance between the cross-sectional mean and the true template.}
    \label{tab: SIM_one cumulative cross-sectional variance}
\end{table}

\subsection{Scenario 2, multivariate functions}\label{sec: multivariate functions}
In this example, the functions to be aligned are $f_i$ where the image of the functions at each $t$ is $\mathbb{R}^3$. We write $f_i(t)=\left(f_{i1}(t),f_{i2}(t),f_{i3}(t)\right)$, and we think of each function $f_{ij}$, as a univariate function that is the observed version of $y_{ij}$ generated as follows. Let $z_{ip}\sim\mathcal{N}(1, 0.25^2)$, $p=1,\dots,18$. The functions $y_{i1}\colon[0,6\pi]\rightarrow\mathbb{R}$ were generated as
\begin{equation*}
    y_{i1}(t)=
    \begin{cases}
        z_{i1}\sin(t)&t\in[0,2\pi),\\
        z_{i2}\sin(t)&t\in[2\pi,4\pi),\\
        z_{i3}\sin(t)&t\in[4\pi,6\pi].\\
    \end{cases}
\end{equation*}
The common template is then $\mu_{1}(t)=\sin(t)$, $t\in[0,2\pi]$. The functions $y_{i2}$ were generated as
\begin{equation*}
    y_{i2}(t)=\begin{cases}
        z_{i4}\left(c_1(x)-c_3(x)\right)+c_3(x)+z_{i5}\left(c_2(x)-c_4(x)\right)+c_4(x),&t\in[0,6),x=t-3,\\
        z_{i6}\left(c_1(x)-c_3(x)\right)+c_3(x)+z_{i7}\left(c_2(x)-c_4(x)\right)+c_4(x),&t\in[6,12),x=t-9,\\
        z_{i8}\left(c_1(x)-c_3(x)\right)+c_3(x)+z_{i9}\left(c_2(x)-c_4(x)\right)+c_4(x),&t\in[12,18],x=t-15,
    \end{cases}
\end{equation*}
where
\begin{alignat*}{2}
    c_1(x)&=e^{-(x-4.5)^2/2},&\quad\quad
    c_2(x)&=e^{-(x-1.5)^2/2},\\
    c_3(x)&=\frac{e^{-\frac{4.5^2}{2}}-e^{-\frac{1.5^2}{2}}}{6}x+\frac{e^{-\frac{4.5^2}{2}}+e^{-\frac{1.5^2}{2}}}{2},&
    c_4(x)&=\frac{e^{-\frac{1.5^2}{2}}-e^{-\frac{4.5^2}{2}}}{6}x+\frac{e^{-\frac{4.5^2}{2}}+e^{-\frac{1.5^2}{2}}}{2}.
\end{alignat*}
This transformation is applied to ensure that $y_{i2}(0)=y_{i2}(6)=y_{i2}(12)=y_{i2}(18)$, allowing continuity at the limits of the periods. The common template is therefore $\mu_{2}(t) =e^{-(t - 4.5)^2/2}+e^{-(t-1.5)^2/2}$, $t\in[-3,3]$. Finally, functions $y_{i3}$ were generated as
\begin{equation*}
    y_{i3}(t)=\begin{cases}
    z_{i10}p_2(x)-z_{i11}(p_1(x)-p_1(0)),&t\in[0,0.5),x=t\\
    z_{i10}p_2(x)+z_{i12}(p_3(x)-p_3(0)),&t\in[0.5,1),x=t\\
    z_{i13}p_2(x)-z_{i14}(p_1(x)-p_1(0)),&t\in[1,1.5),x=t-1\\
    z_{i13}p_2(x)+z_{i15}(p_3(x)-p_3(0)),&t\in[1.5,2),x=t-1\\
    z_{i16}p_2(x)-z_{i17}(p_1(x)-p_1(0)),&t\in[2,2.5),x=t-2\\
    z_{i16}p_2(x)+z_{i18}(p_3(x)-p_3(0)),&t\in[2.5,3),x=t-2,
    \end{cases}
\end{equation*}
where $p_1$ is the probability density function of $\mathcal{N}(0.25,0.1^2)$, $p_2$ is the probability density function of $\mathcal{N}(0.5,0.15^2)$ and $p_3$ is the probability density function of $\mathcal{N}(0.75,0.1^2)$. The template function is then
\begin{equation*}
    \mu_{3}(t)=\begin{cases}
    p_2(t)-(p_1(t)-p_1(0)),&t\in[0,0.5],\\
    p_2(t)+(p_3(t)-p_3(0)),&t\in[0.5,1],
    \end{cases}
\end{equation*}
In addition, functions $y_{i2}$ and $y_{i3}$ were further transformed with an affine transformation so that the image of the template functions $\mu_2$ and $\mu_3$ is in the interval $[-1,1]$. Finally, the domains of all functions $y_{ij}$, as well as the extensions of the template functions, $\ext^K_{\mathrm{L}^2}\mu_{j}$, were scaled to the interval $[0,1]$, and sampled at $P=193$ equidistant points. The functions were then warped as $f_{ij}(t)=\left(y_{ij}\circ\ext^K_{\Gamma}\gamma_i^l\circ\gamma_i^g\right)(t)$.

The optimized architecture consists of an input layer with 193 nodes and three channels, followed by 14 convolutional layers with a kernel of size 60. The hidden layers have 56 filters, while the output layer has only one filter. The learning rate was chosen to be $9.35\cdot10^{-6}$.

\begin{figure}
    \centering
    \begin{subfigure}[b]{0.33\textwidth}
        \centering
        \includegraphics[width=\textwidth]{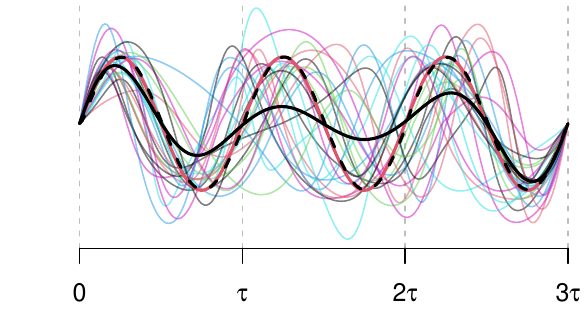}
        \caption{Observed $f_{i1}$}
        \label{fig: SIM three observed 1}
    \end{subfigure}%
    \begin{subfigure}[b]{0.33\textwidth}
        \centering
        \includegraphics[width=\textwidth]{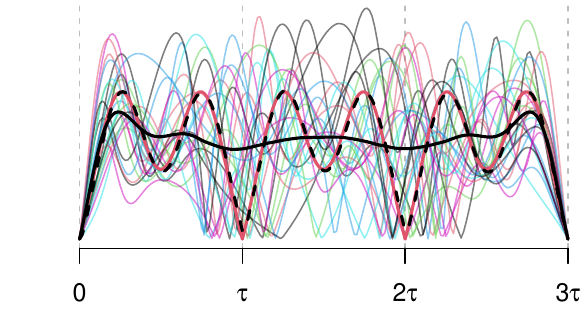}
        \caption{Observed $f_{i2}$}
        \label{fig: SIM three observed 2}
    \end{subfigure}%
    \begin{subfigure}[b]{0.33\textwidth}
        \centering
        \includegraphics[width=\textwidth]{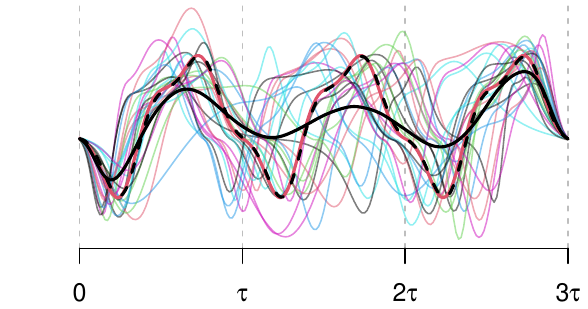}
        \caption{Observed $f_{i3}$}
        \label{fig: SIM three observed 3}
    \end{subfigure}
    \begin{subfigure}[b]{0.33\textwidth}
        \centering
        \includegraphics[width=\textwidth]{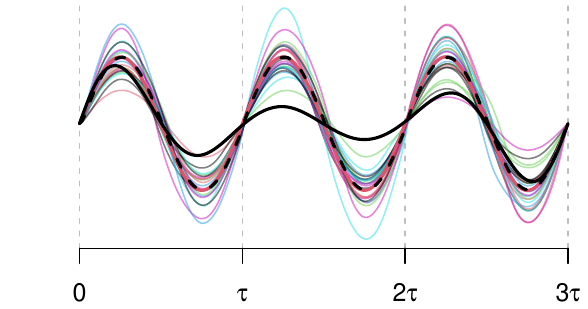}
        \caption{Aligned $f_{i1}\circ\gamma_i$}
        \label{fig: SIM three aligned 1}
    \end{subfigure}%
    \begin{subfigure}[b]{0.33\textwidth}
        \centering
        \includegraphics[width=\textwidth]{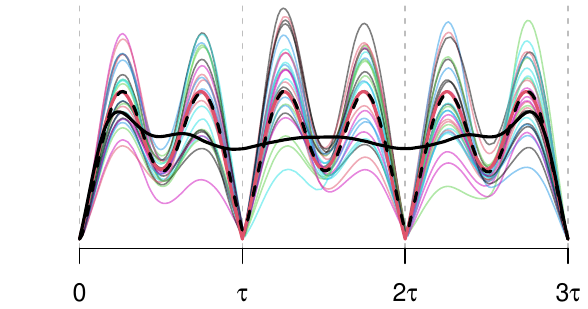}
        \caption{Aligned $f_{i2}\circ\gamma_i$}
        \label{fig: SIM three aligned 2}
    \end{subfigure}%
    \begin{subfigure}[b]{0.33\textwidth}
        \centering
        \includegraphics[width=\textwidth]{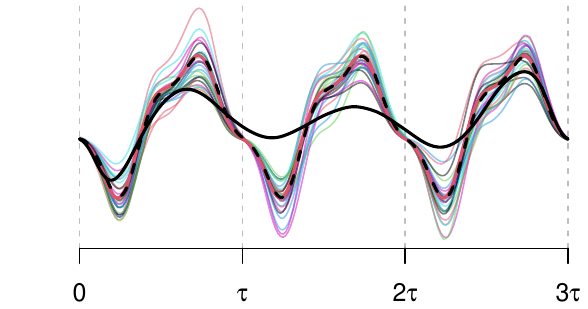}
        \caption{Aligned $f_{i3}\circ\gamma_i$}
        \label{fig: SIM three aligned 3}
    \end{subfigure}
    \caption{Result of the DeepJAM algorithm. Figures \ref{fig: SIM three observed 1}, \ref{fig: SIM three observed 2} and \ref{fig: SIM three observed 3} show random 25 observed functions from the test data. The observed functions are the functions $f_{ij}$, $i=1,\dots,n$, $j=1,2,3$, as described in Section \ref{sec: multivariate functions}. Figures \ref{fig: SIM three aligned 1}, \ref{fig: SIM three aligned 2} and \ref{fig: SIM three aligned 3} show the same functions, but aligned using the DeepJAM neural network. The thick red lines are the true common templates extended to three periods, and the thick black full/dashed lines represent the cross-sectional means of the full observed/aligned test data.}
    \label{fig: SIM three}
\end{figure}

Figure~\ref{fig: SIM three} shows the observed and aligned test functions. We can see that we achieved nearly perfect phase alignment. The functions were simulated with amplitude variability, which is retained in the aligned data. Notice how the cross-sectional means of the observed data misrepresent the intrinsic shape of the underlying data generating model. Furthermore, the periodicity is also lost, especially for the cross-sectional mean of the functions $\{f_{i2}\}$. The cumulative cross-sectional variance of the simulated multivariate data can be found in Table~\ref{tab: SIM_three cumulative cross-sectional variance}.

\begin{table}
    \centering
    \begin{tabular}{rrrrrrrr}
  \toprule
\multicolumn{1}{c}{Functions} & \multicolumn{1}{c}{Observed} & \multicolumn{1}{c}{Aligned} & \multicolumn{1}{c}{$\%\downarrow$} & \multicolumn{1}{c}{Mean} & \multicolumn{1}{c}{Observed} & \multicolumn{1}{c}{Aligned} & \multicolumn{1}{c}{$\%\downarrow$} \\ 
  \midrule
1 & $0.506$ & $0.038$ & $92.54$ & 1 & $0.148$ & $0.001$ & $99.03$ \\ 
  2 & $0.571$ & $0.144$ & $74.76$ & 2 & $0.194$ & $0.003$ & $98.23$ \\ 
  3 & $0.420$ & $0.033$ & $92.06$ & 3 & $0.126$ & $0.002$ & $98.75$ \\ 
   \bottomrule
\end{tabular}

    \caption{Cumulative cross-sectional variance of the observed and aligned simulated multivariate data, where columns ``$\%\downarrow$'' show the reduction of the cumulative cross-sectional variance in $\%$. In addition, we calculate the square of the $\mathrm{L}^2$ distance between the cross-sectional means and the true templates.}
    \label{tab: SIM_three cumulative cross-sectional variance}
\end{table}

\section{Application}\label{sec: application}
For the application, we chose 12-lead 10s ECG recordings. The ECG study population consisted of participants in the Copenhagen General Population Study randomly sampled from the general population in Copenhagen, Denmark \parencite{fuchsSubclinicalCoronaryAtherosclerosis2023, kuhlLeftVentricularHypertrophy2019a}. All ECGs were recorded with the study participant at rest and in supine position prior to a cardiac computed tomography scan. Written informed consent was obtained from all participants, and the study was approved by the local ethics committee (H-KF-01-144/01).
The ECGs were preprocessed using the \texttt{neurokit2} python module \parencite{Makowski2021neurokit}. First, the ECGs were filtered to remove noise and to improve the detection of peaks using the function \texttt{ecg\_clean} with default settings for the method. Afterward, the function \texttt{ecg\_peaks} was used to detect the location of R-peaks, and for the purpose of this application, we only selected a subset of the ECGs between the second and fifth R-peak, giving rise to a three-period time series, with each period between two subsequent R-peaks. The ECGs were then resampled to $P=301$ equidistant points. Furthermore, because of the position of the electrodes, four leads can be derived from other leads \parencite{horacekLeadTheory2010}, thus we will only consider eight leads: I, II, V1--V6. We also only study a subset of the ECGs, resulting in 9,645 ECG recordings. From these, we used 5,511 for training, 1,378 for hyperparameter tuning, 1,378 as the validation data and 1,378 as the test data. 

The optimized architecture consists of an input layer with 301 nodes and eight channels, followed by 17 convolutional layers with a kernel of size 101. The hidden layers have 30 filters, while the output layer has only one filter. The learning rate was chosen to be $2.84\cdot10^{-5}$.

\begin{figure}
    \centering
    \vspace{-2em}
    \begin{subfigure}[b]{0.33\textwidth}
        \centering
        \includegraphics[width=\textwidth]{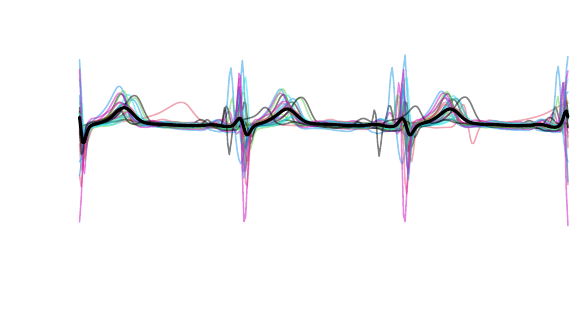}
        \vspace{-3em}
        \caption{Observed lead V3}
        \label{fig: ECG test observed 5}
    \end{subfigure}\hfill
    \begin{subfigure}[b]{0.33\textwidth}
        \centering
        \includegraphics[width=\textwidth]{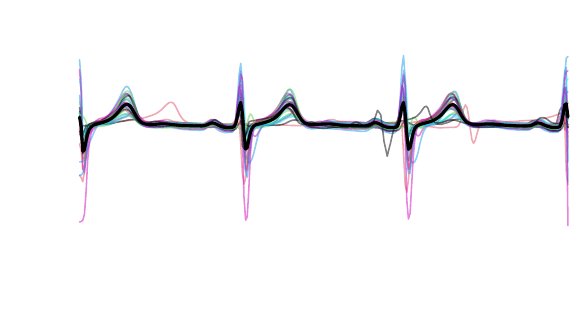}
        \vspace{-3em}
        \caption{Aligned lead V3}
        \label{fig: ECG test aligned 5}
    \end{subfigure}\hfill
    \begin{subfigure}[b]{0.33\textwidth}
        \centering
        \includegraphics[width=\textwidth]{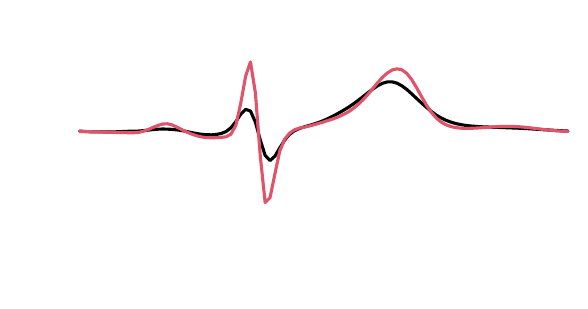}
        \vspace{-3em}
        \caption{Cross-sectional means}
        \label{fig: ECG test means 5}
    \end{subfigure}
    \begin{subfigure}{0.33\textwidth}
        \centering
        \includegraphics[width=\textwidth]{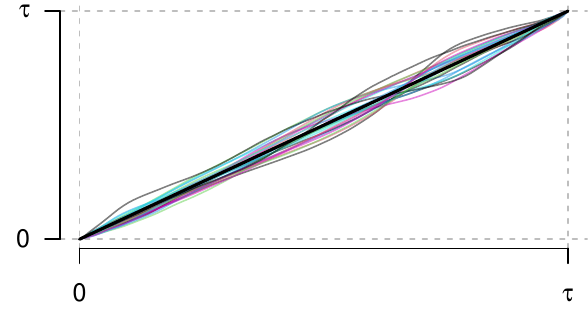}
        \caption{Local warping}
        \label{fig: ECG local warping}
    \end{subfigure}\hfill
    \begin{subfigure}{0.33\textwidth}
        \centering
        \includegraphics[width=\textwidth]{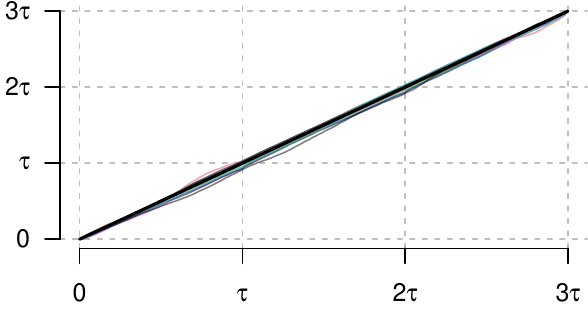}
        \caption{Global warping}
        \label{fig: ECG global warping}
    \end{subfigure}\hfill
    \begin{subfigure}{0.33\textwidth}
        \centering
        \includegraphics[width=\textwidth]{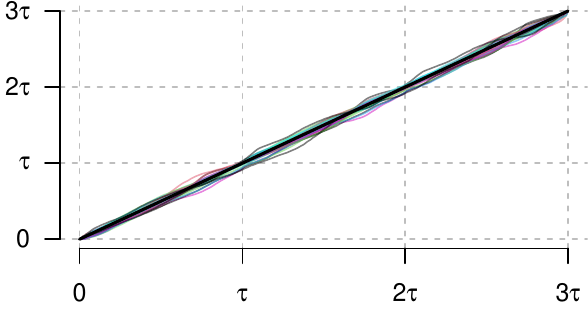}
        \caption{Total warping}
        \label{fig: ECG total warping}
    \end{subfigure}
    \caption{Result of the DeepJAM algorithm. Figures \ref{fig: ECG test observed 5} and \ref{fig: ECG test aligned 5} show the observed and aligned lead V3, respectively, of random 25 test ECGs, where the thick black lines represent the cross-sectional means of the full observed and aligned data, respectively. Figure \ref{fig: ECG test means 5} shows again the cross-sectional means of the full observed data (black) and the full aligned data (red), where we only present one period. Figures \ref{fig: ECG local warping}, \ref{fig: ECG global warping} and \ref{fig: ECG total warping} show the local, global, and total warping of the 25 random test ECGs, with the thick black lines representing the identity line.}
    \label{fig: ECG lead V3}
\end{figure}

Figure~\ref{fig: ECG lead V3} shows the observed and aligned ECG recordings, as well as the corresponding estimated local and global warping functions. In addition, we present a comparison of the cross-sectional means of the observed and aligned data. Table~\ref{tab: ECG cumulative cross-sectional variance} shows the cumulative cross-sectional variance of the observed and aligned data per lead. We can see that alignment using the DeepJAM neural network reduced the cumulative cross-sectional variance by 27.46\%--46.32\%.

\begin{table}
\centering
\begin{tabular}{lrrr}
  \toprule
\multicolumn{1}{c}{Lead} & \multicolumn{1}{c}{Observed} & \multicolumn{1}{c}{Aligned} & \multicolumn{1}{c}{$\%\downarrow$} \\ 
  \midrule
I & $0.005$ & $0.003$ & $32.87$ \\ 
  II & $0.017$ & $0.010$ & $41.08$ \\ 
  V1 & $0.010$ & $0.007$ & $27.46$ \\ 
  V2 & $0.024$ & $0.017$ & $29.50$ \\ 
  V3 & $0.029$ & $0.020$ & $30.68$ \\ 
  V4 & $0.026$ & $0.015$ & $41.23$ \\ 
  V5 & $0.020$ & $0.011$ & $45.57$ \\ 
  V6 & $0.014$ & $0.007$ & $46.32$ \\ 
   \bottomrule
\end{tabular}

\caption{Cumulative cross-sectional variance of the observed and aligned ECG data, where column ``$\%\downarrow$'' shows the reduction of the cumulative cross-sectional variance in~$\%$.}
\label{tab: ECG cumulative cross-sectional variance}
\end{table}

\section{Conclusion}\label{sec: discussion and conclusion}
In this paper, we have presented a novel method for joint alignment of multivariate quasi-periodic functions, addressing the phase variability in the data that is often overlooked by traditional approaches. We equipped our convolutional neural networks with a unique activation function based on the unit simplex transformation, achieving effective phase-amplitude separation. To extend our approach to multivariate functions, we incorporated a convolutional layer with multiple channels in the input layer. The output layer, on the other hand, provides a single warping function per subject, which simultaneously aligns all dimensions of the observed function. We also incorporated the multiscale warping model to accommodate quasi-periodic functions. To train our model, we used a loss function that calculates the Fisher-Rao distance between the square-root slope representations of functions. Notably, our approach is unsupervised and capable of generating common and subject-specific templates.

We conducted experiments on two simulated datasets as well as a real dataset of 12-lead 10s electrocardiogram recordings. We demonstrated the ability of our approach to accurately separate the phase variability from the amplitude variability, with nearly perfect phase alignment in the simulated data and substantial reduction in cumulative cross-sectional variance in the real dataset. We only used very simple convolutional neural networks as a proof of concept. Future work could focus on exploring potential enhancement to the neural network architecture, such as including dropout layers, batch normalization layers, skip layers, as well as different optimizers and activation functions of the hidden layers.

The contribution of our research extends beyond the specific dataset analyzed in this paper. The joint alignment of functions has extensive applications in various fields such as signal processing and computer vision, and medical applications such as neuroscience and data from wearable devices.

\printbibliography
\newpage
\appendix
\section{Observed and aligned ECG test data}
\begin{figure}[h]
    \centering
    \vspace{-1em}
    \begin{subfigure}{0.33\textwidth}
        \begin{subfigure}{\textwidth}
            \centering
            \vspace{-0.5em}
            \includegraphics[width=\textwidth]{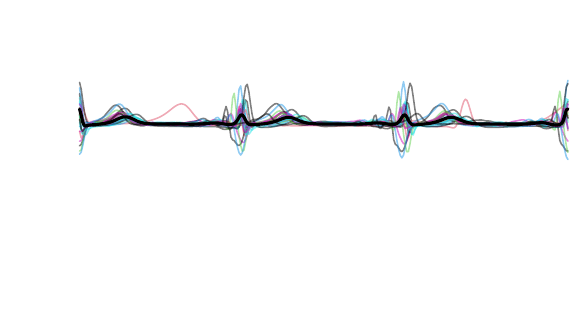}
            \vspace{-4.15em}
            \captionsetup{labelformat=empty}
            \caption{Lead I}
        \end{subfigure}
        \begin{subfigure}{\textwidth}
            \centering
            \vspace{-0.5em}
            \includegraphics[width=\textwidth]{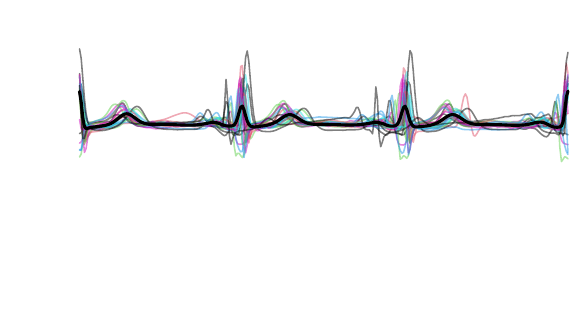}
            \vspace{-4.15em}
            \captionsetup{labelformat=empty}
            \caption{Lead II}
        \end{subfigure}
        \begin{subfigure}{\textwidth}
            \centering
            \vspace{-0.5em}
            \includegraphics[width=\textwidth]{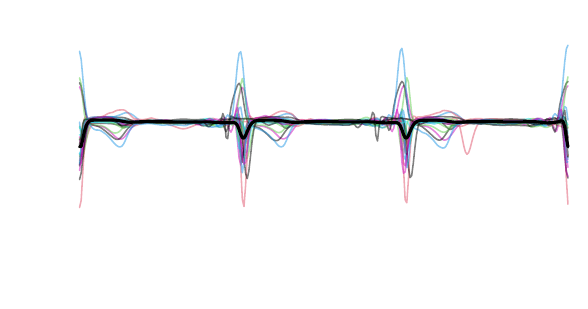}
            \vspace{-4.15em}
            \captionsetup{labelformat=empty}
            \caption{Lead V1}
        \end{subfigure}
        \begin{subfigure}{\textwidth}
            \centering
            \vspace{-0.5em}
            \includegraphics[width=\textwidth]{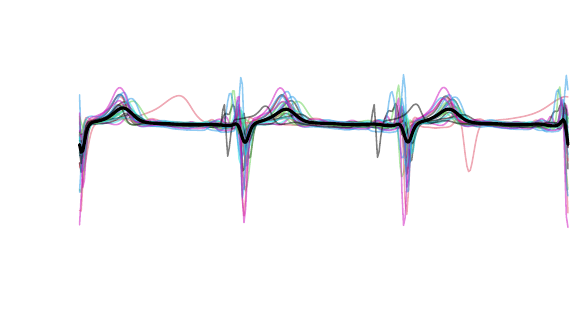}
            \vspace{-4.15em}
            \captionsetup{labelformat=empty}
            \caption{Lead V2}
        \end{subfigure}
        \begin{subfigure}{\textwidth}
            \centering
            \vspace{-0.5em}
            \includegraphics[width=\textwidth]{Figures/ECG_test_observed_5.pdf}
            \vspace{-4.15em}
            \captionsetup{labelformat=empty}
            \caption{Lead V3}
        \end{subfigure}
        \begin{subfigure}{\textwidth}
            \centering
            \vspace{-0.5em}
            \includegraphics[width=\textwidth]{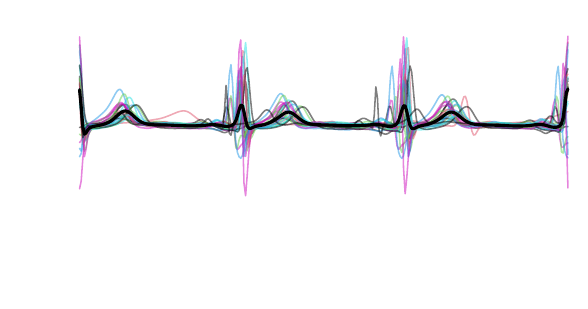}
            \vspace{-4.15em}
            \captionsetup{labelformat=empty}
            \caption{Lead V4}
        \end{subfigure}
        \begin{subfigure}{\textwidth}
            \centering
            \vspace{-0.5em}
            \includegraphics[width=\textwidth]{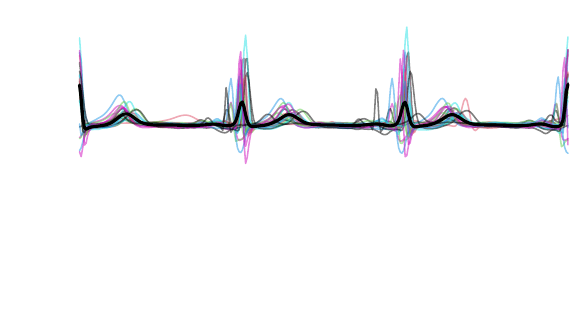}
            \vspace{-4.15em}
            \captionsetup{labelformat=empty}
            \caption{Lead V5}
        \end{subfigure}
        \begin{subfigure}{\textwidth}
            \centering
            \vspace{-0.5em}
            \includegraphics[width=\textwidth]{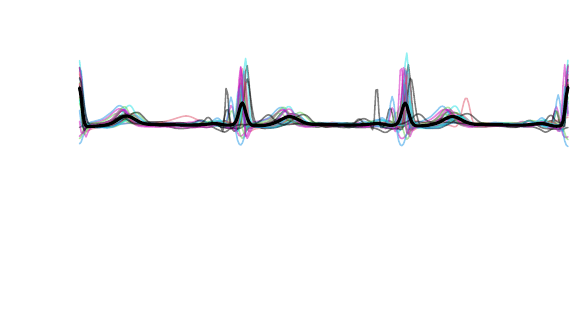}
            \vspace{-4.15em}
            \captionsetup{labelformat=empty}
            \caption{Lead V6}
        \end{subfigure}
        \setcounter{subfigure}{0}%
        \caption{Observed ECGs}
    \end{subfigure}\hfill
    \begin{subfigure}{0.33\textwidth}
        \begin{subfigure}{\textwidth}
            \centering
            \vspace{-0.5em}
            \includegraphics[width=\textwidth]{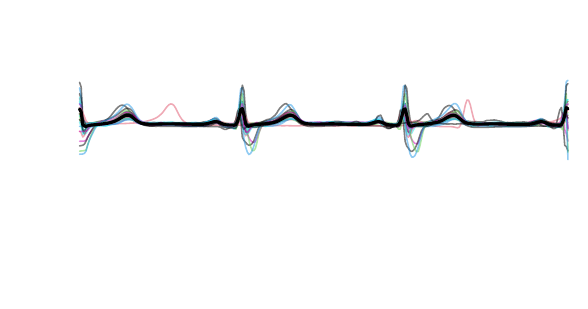}
            \vspace{-4.15em}
            \captionsetup{labelformat=empty}
            \caption{Lead I}
        \end{subfigure}
        \begin{subfigure}{\textwidth}
            \centering
            \vspace{-0.5em}
            \includegraphics[width=\textwidth]{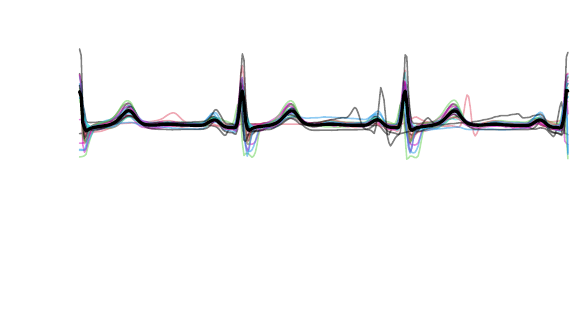}
            \vspace{-4.15em}
            \captionsetup{labelformat=empty}
            \caption{Lead II}
        \end{subfigure}
        \begin{subfigure}{\textwidth}
            \centering
            \vspace{-0.5em}
            \includegraphics[width=\textwidth]{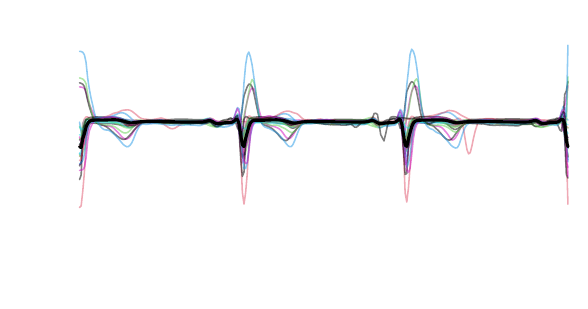}
            \vspace{-4.15em}
            \captionsetup{labelformat=empty}
            \caption{Lead V1}
        \end{subfigure}
        \begin{subfigure}{\textwidth}
            \centering
            \vspace{-0.5em}
            \includegraphics[width=\textwidth]{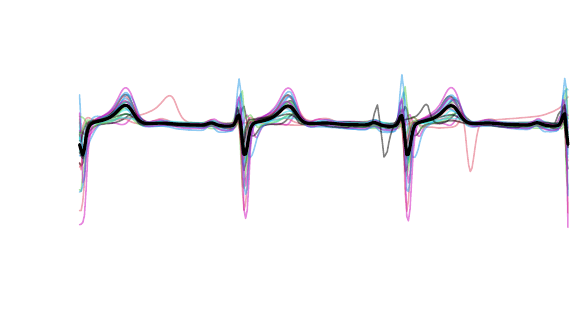}
            \vspace{-4.15em}
            \captionsetup{labelformat=empty}
            \caption{Lead V2}
        \end{subfigure}
        \begin{subfigure}{\textwidth}
            \centering
            \vspace{-0.5em}
            \includegraphics[width=\textwidth]{Figures/ECG_test_aligned_5.pdf}
            \vspace{-4.15em}
            \captionsetup{labelformat=empty}
            \caption{Lead V3}
        \end{subfigure}
        \begin{subfigure}{\textwidth}
            \centering
            \vspace{-0.5em}
            \includegraphics[width=\textwidth]{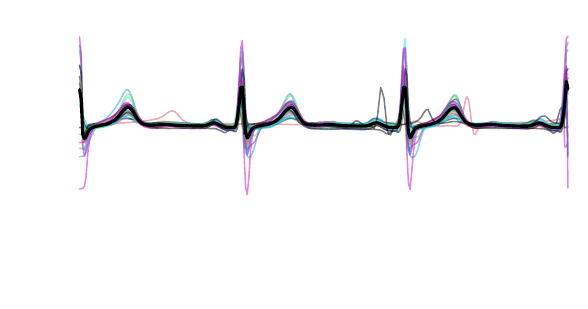}
            \vspace{-4.15em}
            \captionsetup{labelformat=empty}
            \caption{Lead V4}
        \end{subfigure}
        \begin{subfigure}{\textwidth}
            \centering
            \vspace{-0.5em}
            \includegraphics[width=\textwidth]{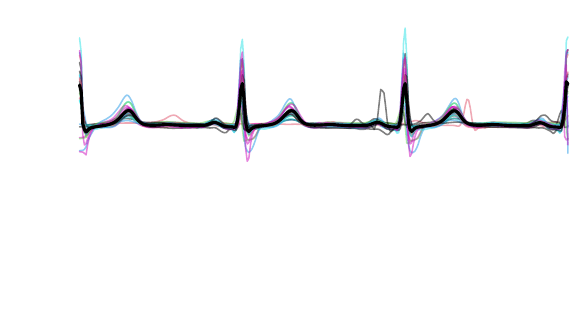}
            \vspace{-4.15em}
            \captionsetup{labelformat=empty}
            \caption{Lead V5}
        \end{subfigure}
        \begin{subfigure}{\textwidth}
            \centering
            \vspace{-0.5em}
            \includegraphics[width=\textwidth]{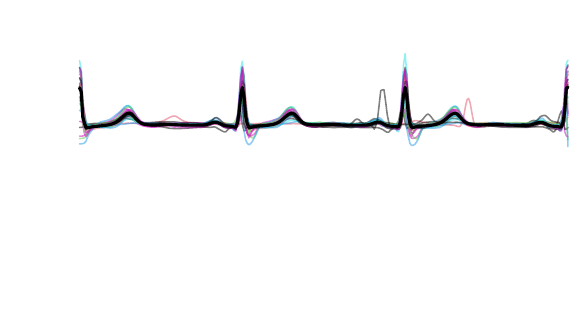}
            \vspace{-4.15em}
            \captionsetup{labelformat=empty}
            \caption{Lead V6}
        \end{subfigure}
        \setcounter{subfigure}{1}%
        \caption{Aligned ECGs}
    \end{subfigure}\hfill
    \begin{subfigure}{0.33\textwidth}
        \begin{subfigure}{\textwidth}
            \centering
            \vspace{-0.5em}
            \includegraphics[width=\textwidth]{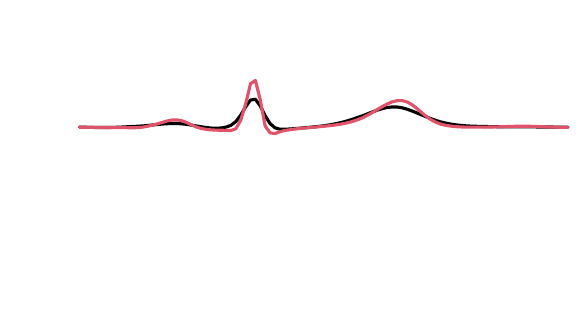}
            \vspace{-4.15em}
            \captionsetup{labelformat=empty}
            \caption{Lead I}
        \end{subfigure}
        \begin{subfigure}{\textwidth}
            \centering
            \vspace{-0.5em}
            \includegraphics[width=\textwidth]{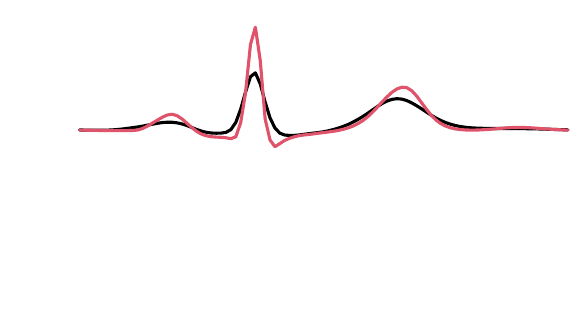}
            \vspace{-4.15em}
            \captionsetup{labelformat=empty}
            \caption{Lead II}
        \end{subfigure}
        \begin{subfigure}{\textwidth}
            \centering
            \vspace{-0.5em}
            \includegraphics[width=\textwidth]{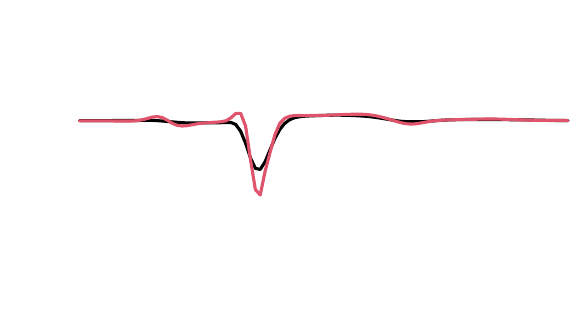}
            \vspace{-4.15em}
            \captionsetup{labelformat=empty}
            \caption{Lead V1}
        \end{subfigure}
        \begin{subfigure}{\textwidth}
            \centering
            \vspace{-0.5em}
            \includegraphics[width=\textwidth]{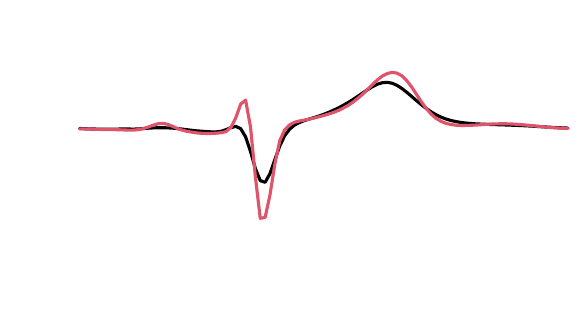}
            \vspace{-4.15em}
            \captionsetup{labelformat=empty}
            \caption{Lead V2}
        \end{subfigure}
        \begin{subfigure}{\textwidth}
            \centering
            \vspace{-0.5em}
            \includegraphics[width=\textwidth]{Figures/ECG_test_means_5.pdf}
            \vspace{-4.15em}
            \captionsetup{labelformat=empty}
            \caption{Lead V3}
        \end{subfigure}
        \begin{subfigure}{\textwidth}
            \centering
            \vspace{-0.5em}
            \includegraphics[width=\textwidth]{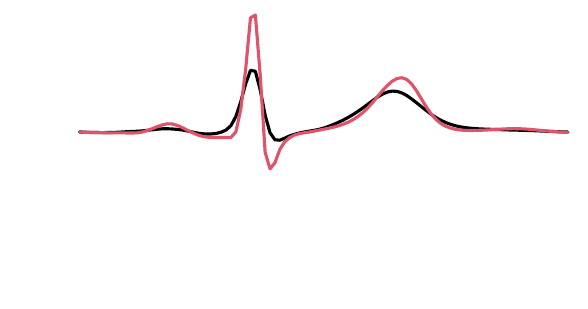}
            \vspace{-4.15em}
            \captionsetup{labelformat=empty}
            \caption{Lead V4}
        \end{subfigure}
        \begin{subfigure}{\textwidth}
            \centering
            \vspace{-0.5em}
            \includegraphics[width=\textwidth]{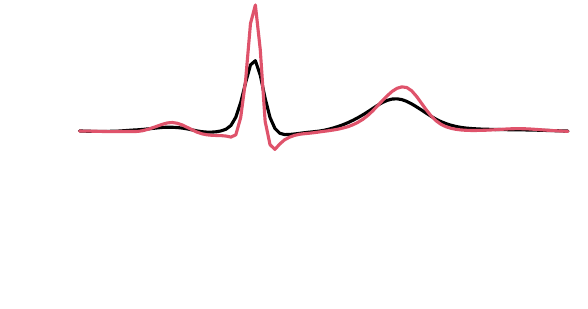}
            \vspace{-4.15em}
            \captionsetup{labelformat=empty}
            \caption{Lead V5}
        \end{subfigure}
        \begin{subfigure}{\textwidth}
            \centering
            \vspace{-0.5em}
            \includegraphics[width=\textwidth]{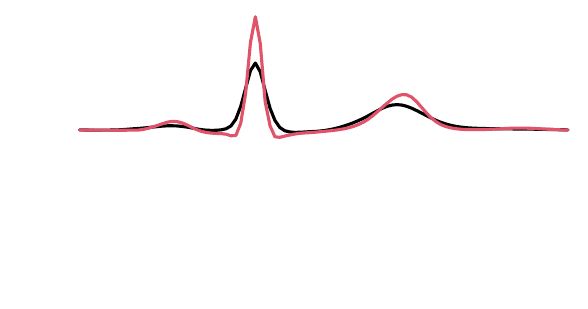}
            \vspace{-4.15em}
            \captionsetup{labelformat=empty}
            \caption{Lead V6}
        \end{subfigure}
        \setcounter{subfigure}{2}%
        \caption{Cross-sectional means}
    \end{subfigure}
    \caption{Result of the DeepJAM algorithm. The first column shows individual leads of random 25 ECG measurements, while the second column shows the same ECGs, but aligned. Finally, the third column shows the cross-sectional means of the full observed dataset (black) and the full aligned dataset (red), where we only show one period.}
    \label{fig: ECG all leads}
\end{figure}

Refer to Figure~\ref{fig: ECG all leads} to see the performance of the DeepJAM algorithm on 8 leads of the ECGs.
\end{document}